# Shear banding in monodisperse polymer melt

Fan Peng, Renkuan Cao, Cui Nie, Tingyu Xu *, and Liangbin Li*

*National Synchrotron Radiation Laboratory, Anhui Provincial Engineering Laboratory of Advanced Functional Polymer Film, CAS Key Laboratory of Soft Matter Chemistry, University of Science and Technology of China, Hefei, 230026, China*

ABSTRACT:

We performed a series of molecular dynamics simulations on monodisperse polymer melts to investigate the formation of shear banding. Under high shear rates, shear banding occurs, which is accompanied with the entanglement heterogeneity intimately. Interestingly, the same linear relationship between the end-to-end distance $R_{ee}$ and entanglement density $Z$ is observed at homogeneous flow before the onset of shear banding and at shear banding state, where $R_{ee} \sim \left[ln(W_i^{0.87}) - \xi_0\right]Z$ is proposed as the criterion to describe the dynamic force balance of molecular chain in flow with a high rate. We establish a scaling relation between the disentanglement rate $V_d$ and Weissenberg number $W_i$ as $V_d \sim W_i^{0.87}$ for stable flow in homogeneous shear and shear banding states. Deviating from this relation leads to force imbalance and results in the emergence of shear banding. The formation of shear banding prevents chain from further stretching and disentanglement. The transition from homogeneous shear to shear banding partially dissipates the increased free energy from shear and reduces the free energy of the system.

Corresponding authors, E-mail: tyxu@mail.ustc.edu.cn (T. Y. Xu); lbli@ustc.edu.cn (L. B. Li)

**For Table of Contents use only**

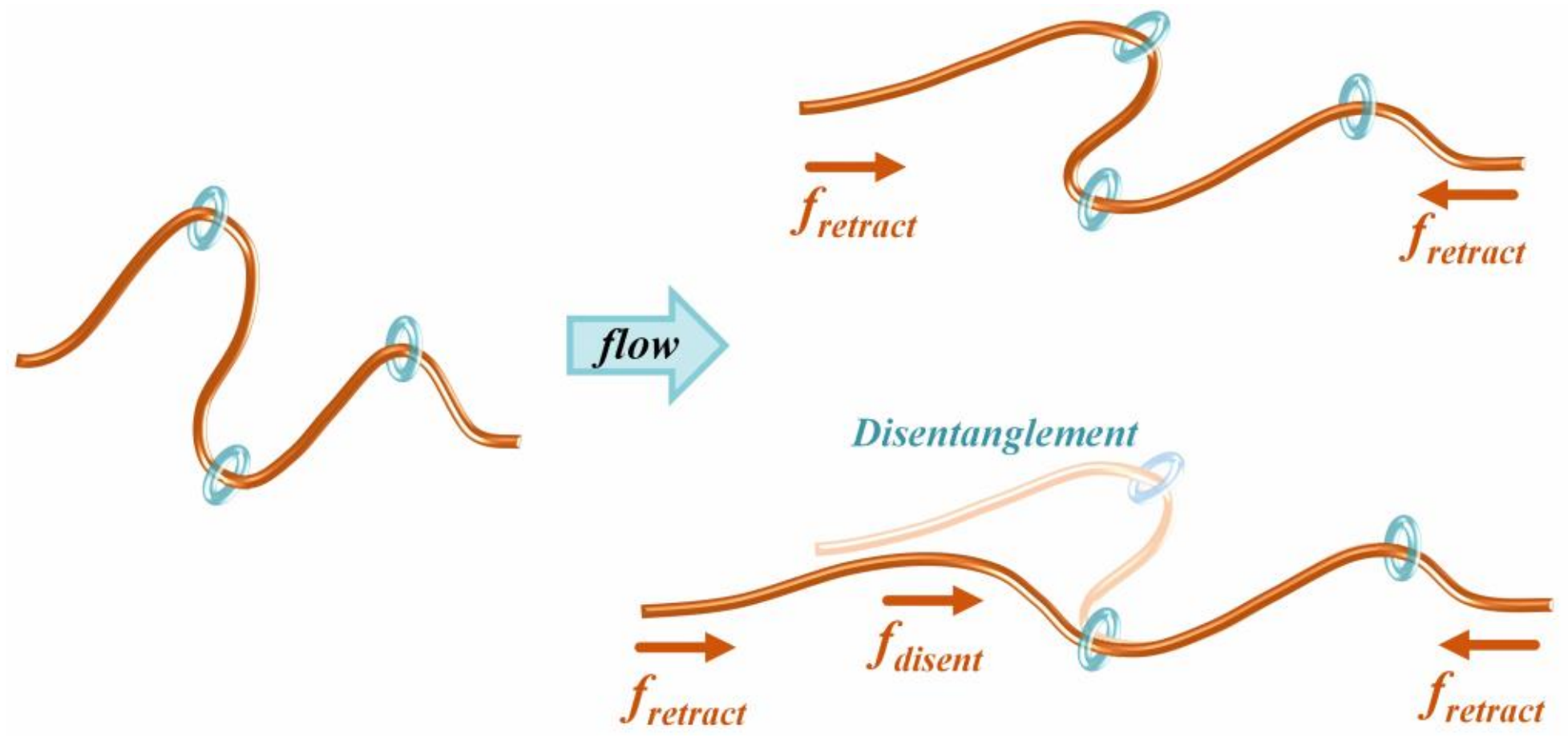

## I. Introduction

Shear banding is a widely observed phenomenon in geological rocks, colloidal solutions, amorphous metals, soft glasses, and polymer systems.[1-8] For polymer solutions or melts, shear banding is a localized yield phenomenon during continuous deformation, indicating non-uniform deformation within the system. Shear banding in polymer solutions was first observed by Callaghan and Gil.[9, 10] With particle tracking velocimetric (PTV) measurements, Wang's group found steady-state shear banding in highly entangled polybutadiene solutions [11] and DNA solutions,[12-14]and shear banding was also observed by others groups,[15-18] providing experimental evidences for the occurrence of shear banding in polymer systems. Shear banding also occurs in industry processing such as extrusion, which affects the deformation of polymer materials.[19-21] If shear banding exists as an intrinsic flow state of polymer,[22-27] the origin of shear banding is of great significance for rheology theory and industry processing, which, however, is still controversial.

Different mechanisms have been proposed to interpret shear banding, including mechanical instability and phase separation. The original tube model supports shear banding stems from non-monotonic constitutive curves.[28, 29] The imbalance of forces in polymer entanglement network was proposed to describe some typical nonlinear rheological behaviors and to provide conditions for the occurrence of the accompanying entanglement heterogeneity of shear banding.[30-33] In polymer solutions, continuous stress-enhanced concentration fluctuation was suggested to result in mechanical shear banding, [34-42] which supports that shear banding can arise from the monotonic constitutive.[43] Burroughs et al. reported shear banding in an entangled polybutadiene solution,[18] which verifies that shear banding is accompanied by non-local flow concentration coupling. Meanwhile, the generalized Gibbs free energy was proposed to describe shear banding. Based on mechanical stress-concentration coupling, Peterson et al. replaced free energy with Lyapunov functional at the condition of constant shear stress,[44] shear banding is seen as a natural consequence of Lyapunov functional minimization. Additionally, phase separation, as a mechanism of stress-induced new phase formation,[45-49] can contribute to shear banding involved in thermodynamic driving. Note that mechanical instability and phase separation are not necessarily contradictory on understanding shear banding, but from either a hydrodynamics or thermodynamics perspective.

Chain entanglement is the most peculiar structure determining not only polymer rheology [50-53] but also other physical behaviors like crystallization, glass transition of polymers.[54-56] In polymer melt, shear banding has been reported to accompany with the heterogeneous distribution of entanglement density.[57-60] Analogue to stress-concentration coupling in polymer solution, non-uniform stress distribution can be generated due to the stress-entanglement density coupling. Shear banding occurs when the chain deforms unevenly under non-uniform stress. Nevertheless, the mechanical and thermodynamic instabilities during shear banding formation in monodisperse polymer melt are still poorly understood.

Molecular dynamics (MD) simulations can give structure changes on molecular scale, wherein the dissipative particle dynamics (DPD) simulations inherently preserve multibody fluid dynamics by maintaining mass and momentum both locally and globally.[61] Using DPD simulation can reproduce some nonlinear rheological phenomena such as shear banding,[62, 63] and present molecular details. To unveil the correlation between entanglement concentration and shear banding, the present work employs DPD simulation to study the formation process of shear banding in monodisperse melt. Shear banding instability is reproduced under different shear rates, and entanglement concentration fluctuation is the necessary pre-condition for the occurrence of shear banding. A correlation between entanglement concentration $Z$ and end-to-end distance $R_{ee}$ can be taken as the criterion for dynamic force balance during shearing. Moreover, the occurrence of shear banding dissipates part of the free energy stored in system, and shear banding can also be understood with thermodynamic argument.

## 2. Simulation Method

All simulations are performed on open-source code LAMMPS.[64] The Kremer−Grest model with bending potential is used.[65, 66] The non-bonding interactions are given by a shifted purely repulsive pairwise 12-6 Lennard-Jones (LJ) potential $U_{nonb} = 4\varepsilon\left[\left(\frac{\sigma}{r_{ij}}\right)^{12} - \left(\frac{\sigma}{r_{ij}}\right)^{6} + \frac{1}{4}\right]\Theta\left(r_c - r_{ij}\right)$, where $\sigma$ is the LJ diameter and $r_{ij}$ is the distance between monomer $i$ and $j$. $\Theta(x) = 0$ for $x < 0$, and $\Theta(x) = 1$ for $x \geq 0$. The LJ potential is truncated at $r_c = 1.1225\sigma$. The bonding interactions is given by an infinitely extensible nonlinear elastic (FENE) potential $U_{bond} = -\frac{k}{2}R_0^2 ln[1 - (r/R_0)^2]+4\varepsilon\left[\left(\frac{\sigma}{r_{ij}}\right)^{12} - \left(\frac{\sigma}{r_{ij}}\right)^{6} + \frac{1}{4}\right]\Theta\left(r_c - r_{ij}\right)$, where the spring constant k = $30\varepsilon/\sigma^2$, $R_0$ = 1.5$\sigma$. The chain stiffness is considered by a bending potential $U_{bend} = k_b(1 - cos\theta)$ with $k_b$ = 2$\varepsilon$.

The model of polymer melt contains 1646 chains with chain length $N$ = 300. The monomer density is set as 0.85$\sigma^{-3}$. The disentanglement time $\tau_d$ is about 3.67 × $10^6\tau$.[57] Considering the slow relaxation of the melt, the double bridge algorithm is used to accelerate the relaxation process.[66] The mean square internal distances of molecular chains in initial equilibrium system is plotted in Figure S1 of supporting information. At equilibrium, the root-mean-square radius of gyration $R_{g0}$ is 12.6$\sigma$ and the number of entanglement points on each chain is $Z$ = 28 (the initial average entanglement length is 10.6), which is calculated by Z1+ code.[67-71] The simulation box with $L_x$ = 76.2$\sigma$, $L_y$ = 101.6$\sigma$, $L_z$ = 76.2$\sigma$ and the periodic boundary conditions are set. Then, the polymer melt is subjected to shear flow field with different Weissenberg numbers $W_i$ = 5 ~ 1000, where $W_i = \tau_d\dot{\gamma}$, and $\dot{\gamma}$ is shear rate. The shear flow is along the $x$-axis and produces a velocity gradient in the $y$-axis. During shearing, the reduced temperature $T$ = 1 is controlled by the DPD thermostat,[57, 62] and the simulation time step is chosen as 0.005$\tau$.

## 3. Results

**3.1. The effect of shear rate.** The polymer melt shows different instability behaviors under different shearing rates. Shear banding shows shear rate dependence, which is consistent with previous simulations.[57]

For $W_i$ = 5, no visible shear banding appears. Figure 1a shows the dimensionless velocity profiles at different strains $\gamma$, wherein $V(y)$ is the moving speed of the monomer in the velocity gradient direction $y$, reduced by strain rate $\dot{\gamma}$ and the root-mean-square radius of gyration in equilibrium state $R_{g0}$. The velocity profile remains linear with the change of $\gamma$, indicating that shear banding does not occur. The number of entanglement points $N_z$ in different layers along $y$-axis at different $\gamma$ is plotted in Figure 1b. In the initial system without shear, $N_z$ is unevenly distributed around $N_z$ = 450, but it gradually evolves into a uniform distribution around $N_z$ = 420 with the increase of $\gamma$.

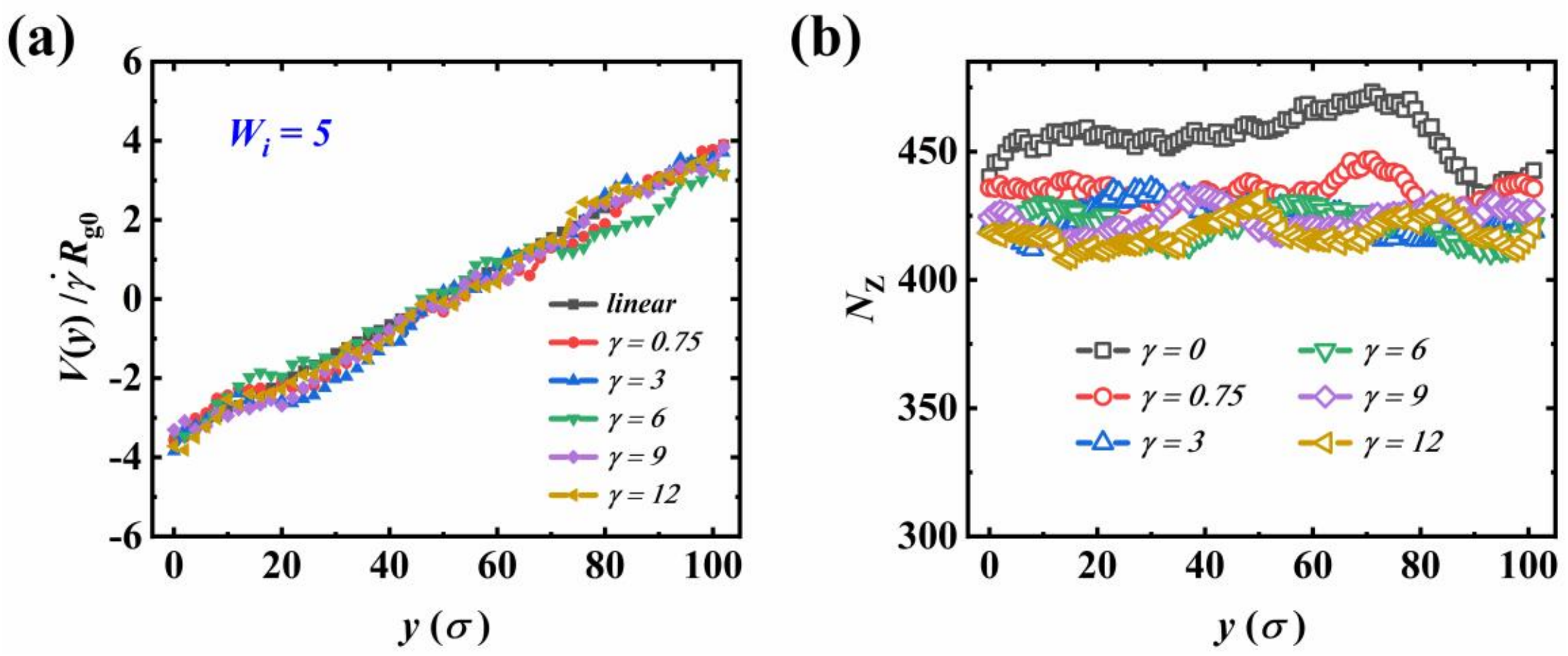

Figure 1. (a) The velocity reduced by strain rate and the root-mean-square radius of gyration in gradient direction. (b) The distribution of the number of entanglement points $N_z$ in gradient direction at different $\gamma$.

For $W_i$ = 50 and 100, flow induces shear banding which is accompanied with heterogeneous entanglement distribution. Figures 2a and 2b give the reduced velocity profiles at different $\gamma$ under $W_i$ = 50 and 100. The velocity profiles remain linear after shear is applied, then suddenly switch to nonlinear at $\gamma$ = 8 and the nonlinearity persists in subsequent shearing. The nonlinear velocity profiles show two bands with sharp boundaries, which are consistent with the characteristics of shear banding. In Figures 2c and 2d, $V'$, the derivative of $V(y)$ with respect to $y$, is presented at $\gamma$ = 24 when the bands is fully formed. The reduced $V'$ sharply fluctuates from 0 to 0.35 at $W_i$ = 50, and from 0 to 0.4 at $W_i$ = 100. The regions with continuous $V'(y)/\dot{\gamma}R_{g0} > 0.05$ are identified as fast band and conversely as slow band. Two bands are observed at $W_i = 50$ and 100, and the locations of bands are not fixed. The ranges of coordinates of slow band are 0 ~ 18 and 50 ~ 102 at $W_i = 50$, and 16 ~ 86 at $W_i = 100$. The widths of slow bands are nearly the same of about 70, always larger than that of fast band. Figures 2e and 2f provide $N_z$ in different layers along $y$-axis at different $\gamma$. $N_z$ fluctuates from 430 to 470 at $\gamma = 0$. At $\gamma = 24$, $N_z$ fluctuates from 340 to 410 at $W_i = 50$, and from 325 to 405 at $W_i = 100$, while the difference between the maximum and minimum varies little with $W_i$. $N_z$ decreases in all layers during shearing, but the decrease is not uniform in $y$-axis and the heterogeneity of $N_z$ distribution gradually enhances as $\gamma$ increases. There is a valley of $N_z$ distribution curve around $y = 90$ at $\gamma = 0$, which gradually moves towards $y = 40$ under $W_i = 50$. Under $W_i = 100$, the valley of $N_z$ distribution curve is $y = 100$ at $\gamma = 24$. $N_z$ shows another minimum around $y = 40$, which however is much larger than that in $y = 100$. The positions of peaks and valleys of $N_z$ distribution curves in Figure 2e/2f coincide with those of the slow and fast bands in Figure 2c/2d, and also correspond to the bands of velocity in Figure 2a/2b, respectively.

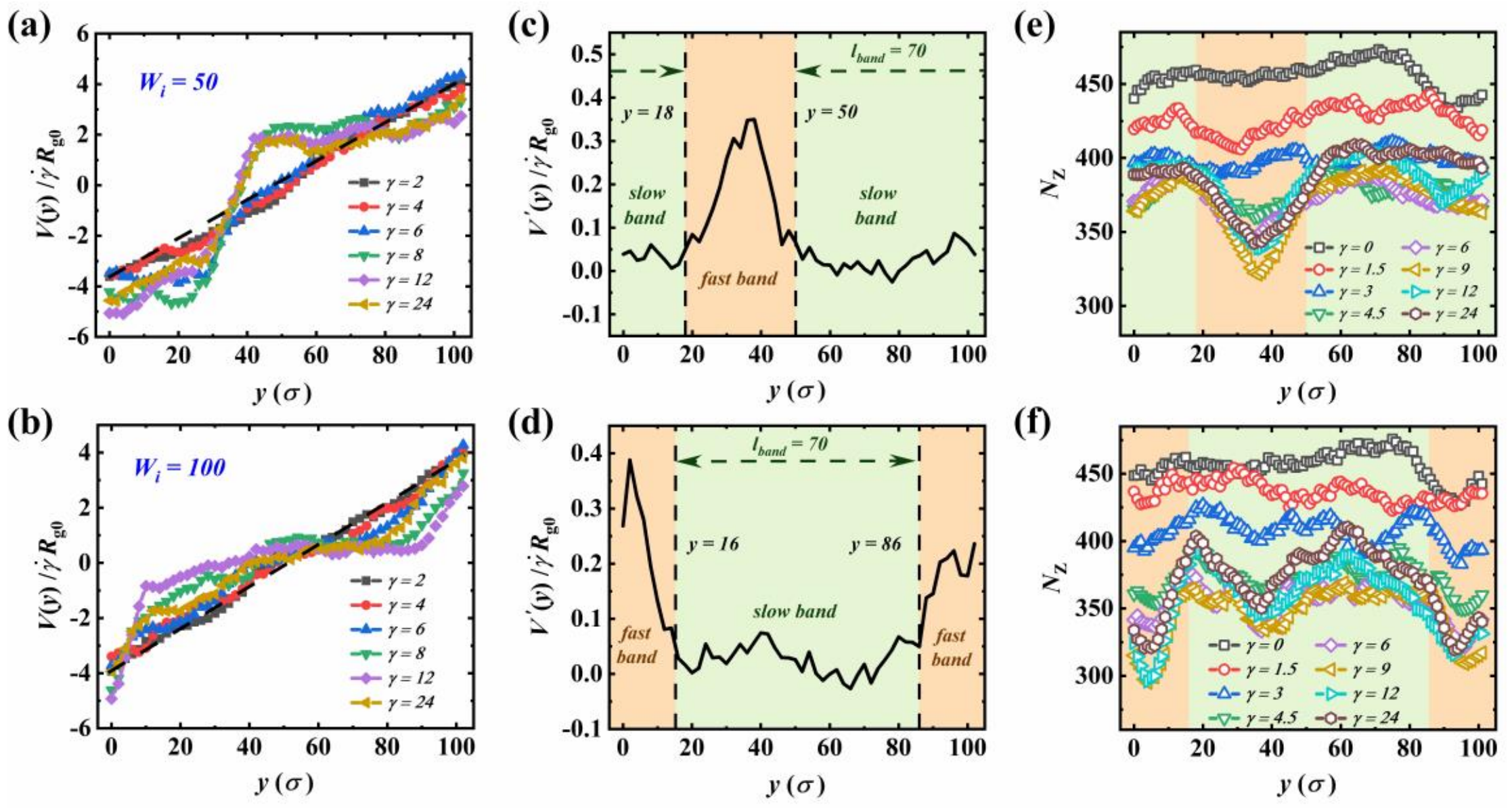

Figure 2 The dimensionless velocity, the derivative of the dimensionless velocity with respect to $y$, and the number of entanglement points in different layers along the velocity gradient direction at different $\gamma$ at $W_i$ = 50 and 100. The black dotted lines in (a) and (b) are linear velocity contrast. The yellow filled areas correspond to the fast band areas, and the green ones correspond to the slow band areas.

Shear banding weakens with a further increasing shear rate. Figures 3a and 3b give the velocity profiles at different $\gamma$ under $W_i$ = 500 and 1000, the velocity distributions gradually deviate from linearity as $\gamma$ increases. The bands at $W_i$ = 500 are relatively blurred and the number of bands is larger than that at smaller $W_i$, but shear banding does occur. The velocity distribution weakly deviates from linearity at $W_i$ = 1000. The reduced $V'$ as a function of $y$ at $\gamma$ = 48 is plotted in Figures 3c and 3d. The reduced $V'$ fluctuates from 0 to 0.2 at $W_i$ = 500, and from 0 to 0.14 at $W_i$ = 1000, in which the amplitude of fluctuation is significantly smaller than that at smaller $W_i$. Choosing the reduced $V'$ of 0.05 as the dividing line between fast and slow bands, there are four bands at $W_i$ = 500 and 1000. The total width of the slow bands is 34 at $W_i$ = 500, and 22 at $W_i$ = 1000, which is much smaller than the width of 70 at $W_i$ = 50/100. Figures 3e and 3f provide $N_z$ distributed in different layers along $y$-axis at different $\gamma$. The fluctuation of $N_z$ distribution enhances as $\gamma$ increases. At $\gamma$ = 48, $N_z$ ranges from 280 to 345 at $W_i$ = 500 and from 240 to 320 at $W_i$ = 1000. The degree of disentanglement significantly increases with $W_i$, and the difference between the maximum and the minimum keeps almost constant at different $W_i$. The coordinates of the peak/valley of $N_z$ distribution strictly correspond to the position of slow/fast bands, respectively, again suggesting strong correlation between shear banding and disentanglement of polymer chains.

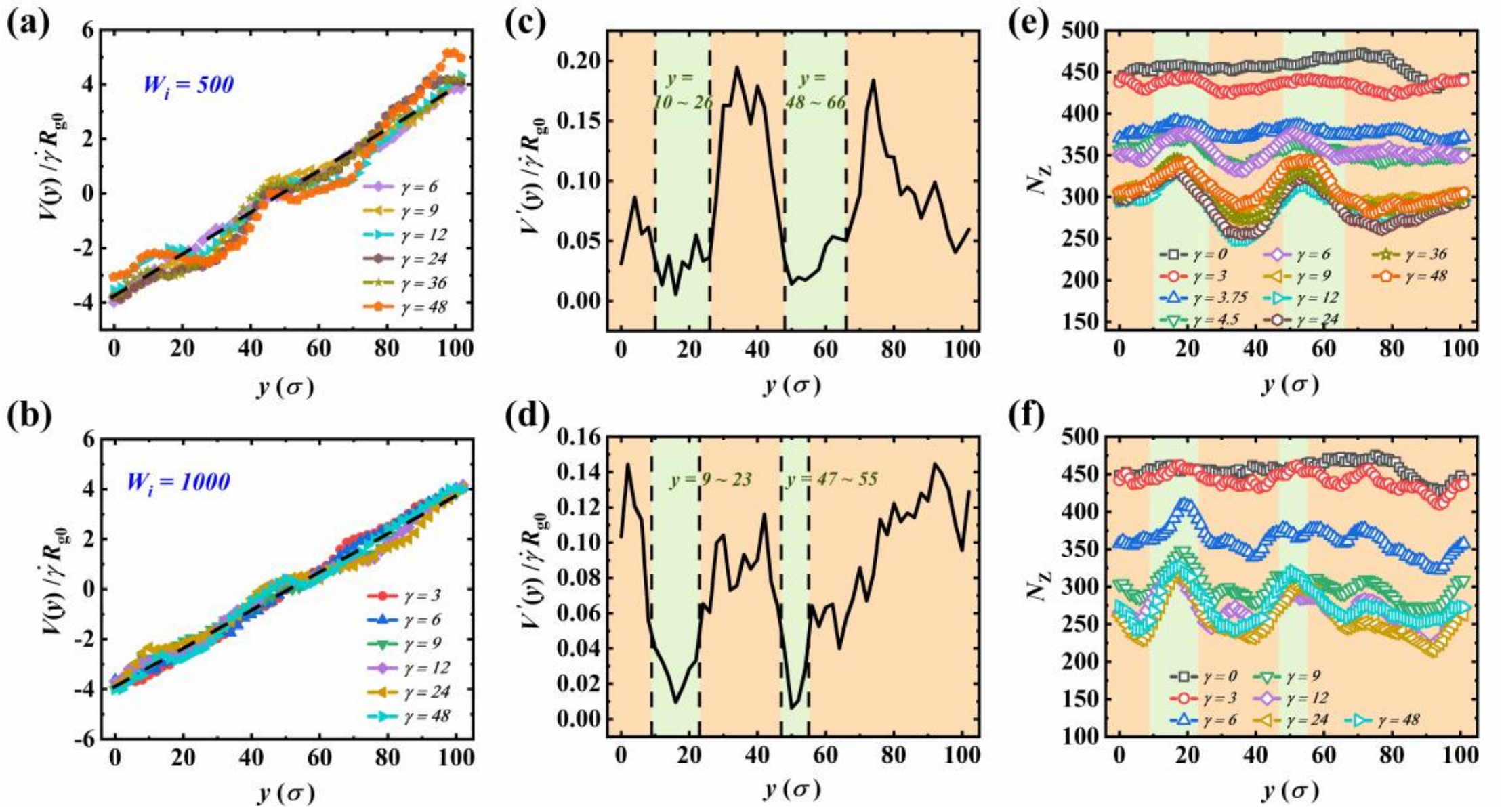

Figure 3 The dimensionless velocity, the derivative of the dimensionless velocity with respect to $y$, and the number of entanglement points in different layers along the velocity gradient direction at different $\gamma$ at $W_i$ = 500 and 1000. The black dotted line in (a) and (b) as linear velocity contrast. The yellow and green filled areas correspond to the fast and slow band areas.

$W_i$ can be divided into two intervals depending on whether shear banding occurs.

Under small $W_i$, shear banding cannot occur and homogeneous entanglement remains. Upon higher $W_i$, shear banding emerges accompanied with heterogeneous entanglement, but the clarity of bands is reduced with the increase of $W_i$. With the increase of $W_i$, the velocity profile evolves from the clear nonlinearity to a weak deviation from linearity, the maximum value of reduced $V'$ decreases and the width of slow band reduces, while the difference between the peaks and valleys of $N_z$ distribution remains in a small variation.

**3.2. The evolution of velocity and conformation during shear.** As shown in Figures 2 and 3, the distribution of the reduced $V'$ correlates with that of $N_z$. To confirm this correlation, we define the difference of the average $V'$ in fast and slow bands as $\Delta V'$, and the difference between the average $N_z$ in fast and slow bands as $\Delta N_z$. Figure 4 depicts $\Delta V'$ and $\Delta N_z$ as the functions of $\gamma$ during shear. Upon starting shear, $\Delta V'$ and $\Delta N_z$ evolve synchronously with a short flat, an increase and a plateau strain regions. Velocity fluctuation is tightly associated with heterogeneous entanglement during the whole shearing process, and the response of velocity to heterogeneous entanglement is instantaneous.

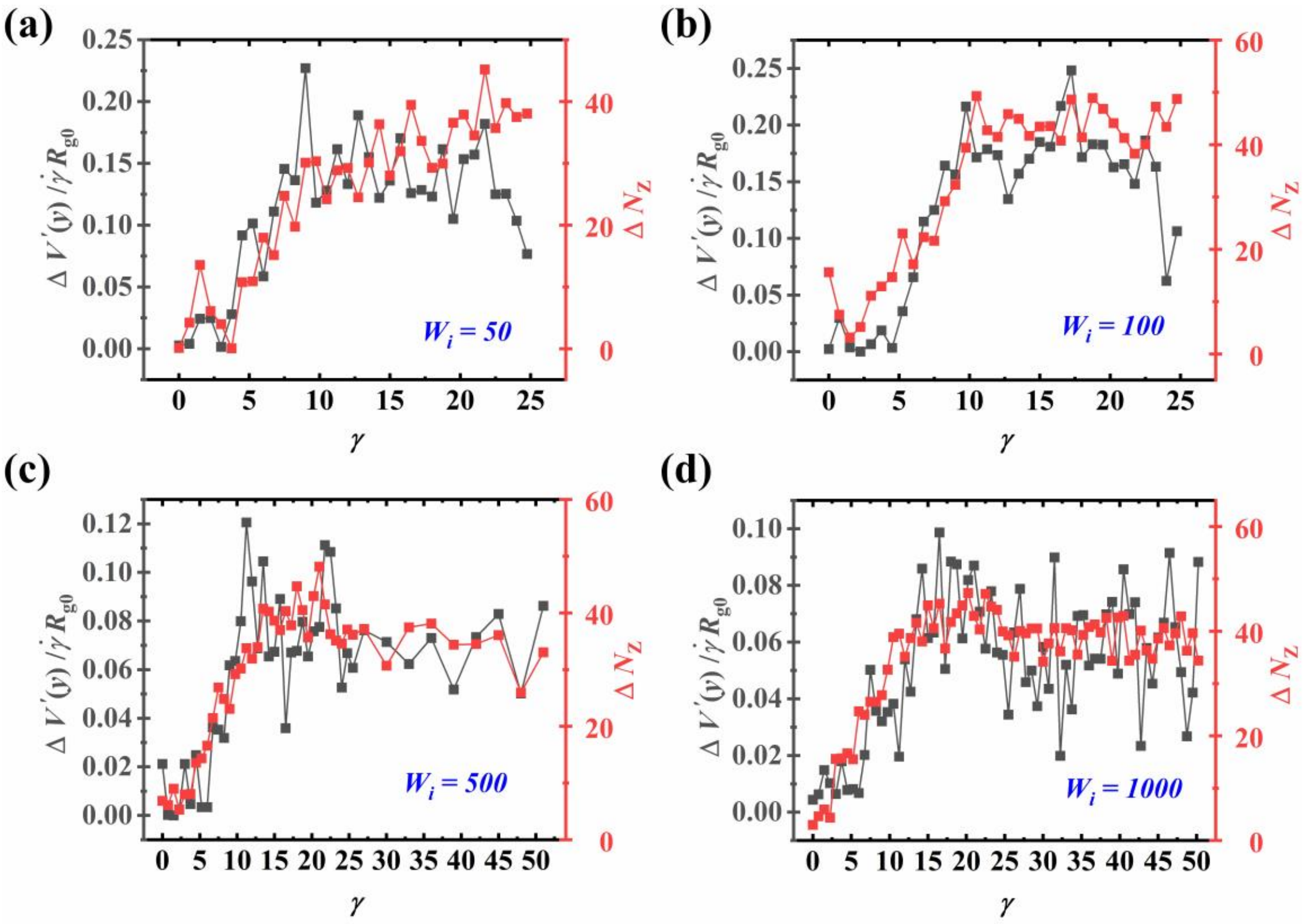

Figure 4 The difference between the average value of $V'$ in fast and slow bands (left coordinate axis) and that of $N_z$ in fast and slow bands (right coordinate axis) as the function of $\gamma$ at different $W_i$.

How do the structural changes at different length scales correlate with shear banding? To display the evolution of the structures at different scales during shearing, the average orientation parameter $P_2$, the average number of entanglement points $\langle N_z \rangle$ in layers, and the average end-to-end distance $\langle R_{ee} \rangle$ are calculated. $P_2 = \langle 3\cos^2\theta_{i,j} - 1 \rangle/2$ represents the local orientational order on monomers scale,[72] where $\theta_{i,j}$ is the angle between two adjacent string vectors. $\langle N_z \rangle$ in strand scale is obtained by dividing the total number of entanglement points in the fast/slow bands by the widths of fast/slow bands in the velocity gradient direction. Figure 5 depicts $P_2$, $\langle N_z \rangle$, and $\langle R_{ee} \rangle$ in the whole system, and the fast/slow bands as the function of $\gamma$ at $W_i$ = 50, 100, and 500. With the increase of $\gamma$, $P_2$ and $\langle R_{ee} \rangle$ increase first and then decreases, while $\langle N_z \rangle$ decreases first and then increases. The black, blue, and red dashed lines mark the $\gamma$ when stress overshoot (see Figure S2 of supporting information for stress-strain curve), the difference of $\langle N_z \rangle$ in the fast and slow band regions emerges, and shear banding is fully formed, respectively. Note here "band regions" represent the corresponding locations of fast and slow bands before shear banding emerges. The stress overshoot, heterogeneous entanglement and shear banding appears in succession. At $W_i$ = 50 and 100, $P_2$ exhibits difference between fast and slow band regions around $\gamma$ = 3, immediately after the occurrence of stress overshoot at $\gamma$ = 2 and 2.3, respectively. The heterogeneities of $\langle N_z \rangle$ and $\langle R_{ee} \rangle$ appear later around $\gamma$ = 4.5 and 5.25, respectively. Under all $W_i$, $P_2$, $\langle N_z \rangle$ and $\langle R_{ee} \rangle$ follow a nonmonotonic or a rebound process with $\gamma$, suggesting chain retraction and entanglement reconstruction. This rebound phenomenon is a necessary but not sufficient condition for shear banding formation, as it is also observed in other short-chain systems where shear banding does not occur (see Figure S3 of supporting information).

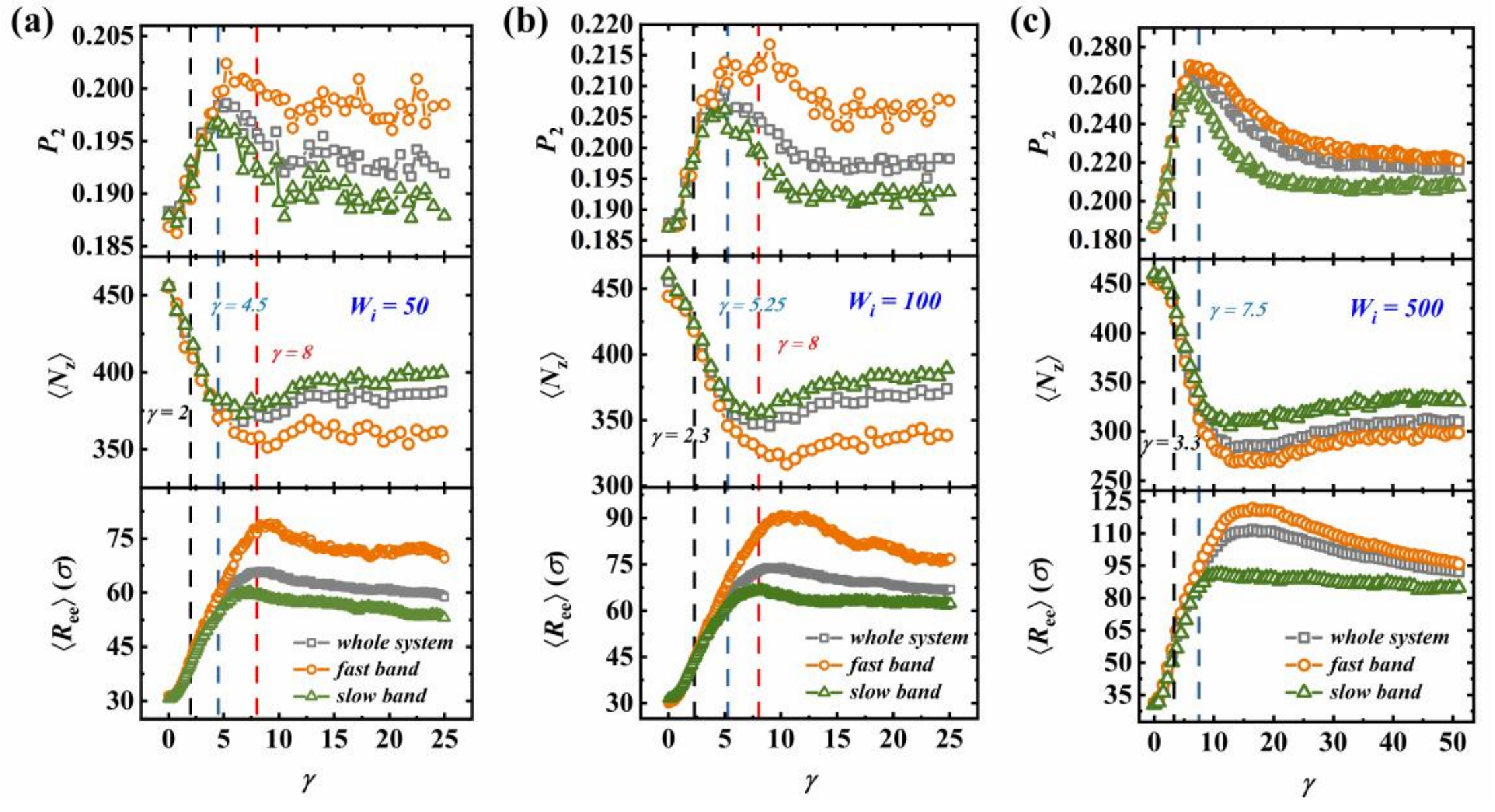

Figure 5 The orientational order parameter $P_2$, the average entanglement points density

$\langle N_z \rangle$ in each layer and the average end-to-end distance $\langle R_{ee} \rangle$ of chain as the function of strain at (a) $W_i$ = 50, (b) $W_i$ = 100, (c) $W_i$ = 500. The grey, yellow and green represent the changes in the whole system, the fast and slow band regions, respectively. The black, blue, red dash lines mark the onset strains of stress overshoot, heterogeneity of $\langle N_z \rangle$ occurs, and shear banding, respectively.

The average $\langle R_{ee} \rangle$ and $\langle N_z \rangle$ evolve synchronously but with opposite trend with $\gamma$ during shear (Figure 5), indicating chain stretching and disentanglement are correlated. Figure 6 provides the end-to-end distance $R_{ee}$ as a function of the number of entanglement points (entanglement density) $Z$ on each chain at different $W_i$. $Z$ decreases from right to left of abscissa, corresponding to that $R_{ee}$ increases gradually as $\gamma$ increases. The arrows between the symbols indicate the direction of the evolution with $\gamma$. Every two adjacent points are separated by strain of 0.75. Under small $W_i$ = 5 where no shear banding emerges, shear induces relatively small change for both $R_{ee}$ and $Z$ as Figures 6a shows, the corresponding monotonic changes of $R_{ee}$ and $Z$ with strain are plotted in Figure S4 of Supporting Information. Except for the shear startup stage, the $Z$-$R_{ee}$ evolves nearly linearly along the blue dashed line. Here the shear startup stage is defined as the shear stage before the stress is close to the overshoot value after shear application (around 0.93 of the stress overshoot value, see Figure S5 and Table S1). Figures 6b and 6c give $Z$-$R_{ee}$ plots under $W_i$ = 50 and 100, respectively. Here the yellow and green solid symbols represent fast and slow bands before the occurrence of shear banding, respectively, while the corresponding hollow symbols represent those after shear banding, where the gradual color fading represents the increase in $\gamma$. The data are connected by arrows along the increase of $\gamma$. The whole shear process can be divided into four stages: shear startup, homogeneous shear before shear banding, the transition stage from homogeneous shear to shear banding, and shear banding state. Ignoring the shear startup stage, $Z$-$R_{ee}$ plots falls nearly in a straight line first and then begins to deviate around $\gamma = 3$, where heterogeneity of $P_2$ occurs. Such deviation remains with increasing inhomogeneity of the system until shear banding occurs. When further increasing strain after attaining shear banding, $Z$-$R_{ee}$ plots gradually returns to the original straight line, as illustrated with the blue dash line in Figures 6b and c. Interestingly, $Z$-$R_{ee}$ plots for both flow states before and after shear banding follow the same straight line (see the blue dash line in Figures 6b and c), while deviation only exists in the flow states during the transition from homogeneous flow to shear banding. In homogeneous shear and shear banding stage, the heterogeneity of orientation and entanglement does not change with strain, these two stages are defined as the stable flow in subsequent discussions. Figure 6 implies that a balance is held between $R_{ee}$ and $Z$ in the stable flow. Accordingly, the imbalance between $R_{ee}$ and $Z$ leads to unstable flow and eventually the occurrence of shear banding.

The same negative linear relation between $R_{ee}$ and $Z$ is observed under $W_i = 500$ and $W_i = 1000$, which is plotted in Figure 6d. In addition to the arrow, the color fading also represents $\gamma$ from small to large. As both $R_{ee}$ and $Z$ rebound back after the emergence of shear banding, leading to difficulties in illustrating the increase of $\gamma$ with arrow alone. Under both $W_i$, $R_{ee}$ changes with $Z$ linearly except the stage of shearing startup, and eventually stabilizes. Comparing with those under $W_i = 50$ and $W_i = 100$, the $Z$-$R_{ee}$ deviation from linearity in the unstable flow state is small, which is possibly due to small difference of $V$ between fast and slow bands. Additionally, the shear of $W_i$ = 25, 75, 250 are performed (see Figure S6-S8 for conformational evolution during shearing), confirming the same linear relationship under different $W_i$.

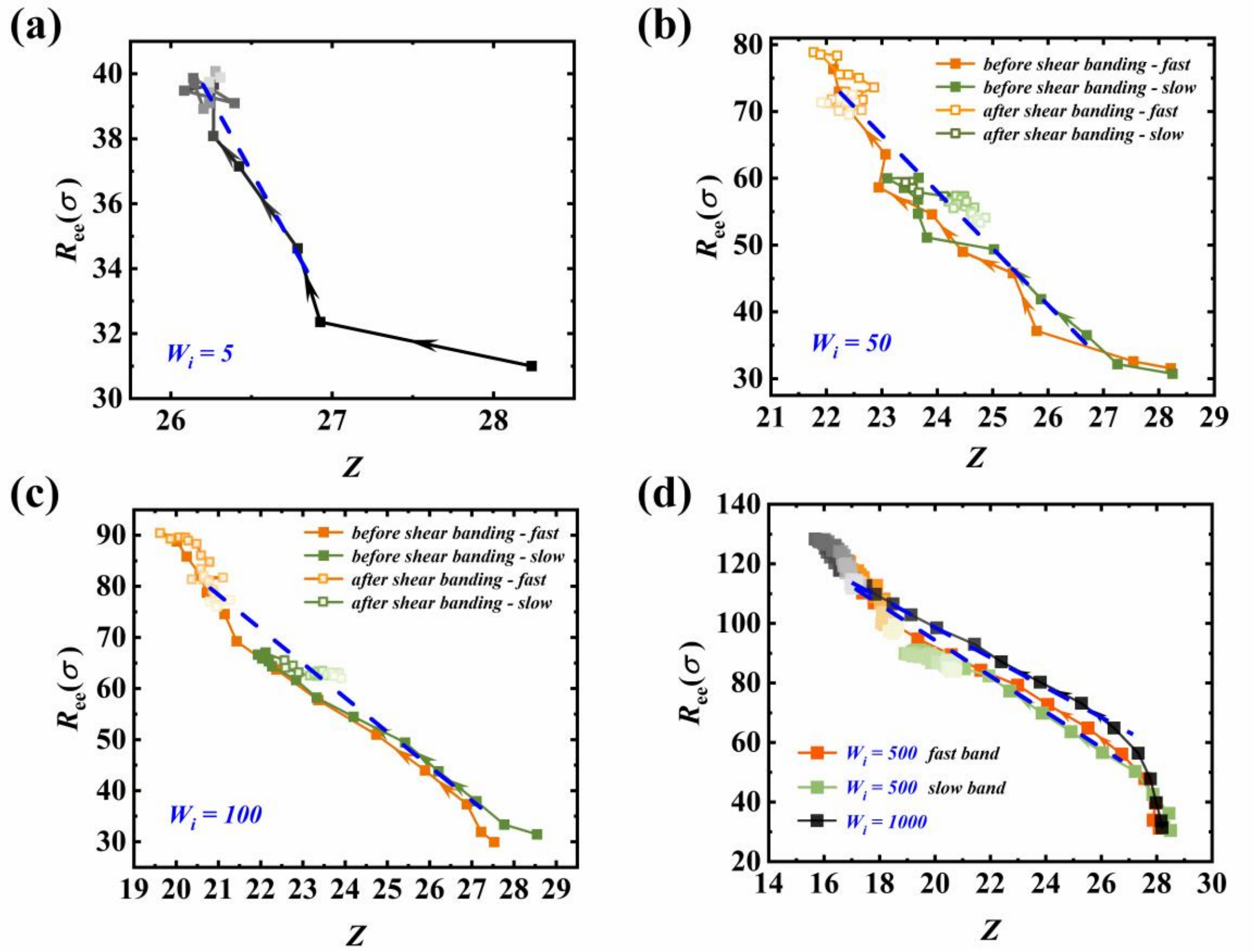

Figure 6 The average end-to-end distance as a function of the average number of entanglement points per chain at (a) $W_i$ = 5, (b) $W_i$ = 50, (c) $W_i$ = 100, (d) $W_i$ = 500 and 1000. The arrows between the symbols indicates the direction of $Z$-$R_{ee}$ changes with $\gamma$ increase. The black symbols is $Z$-$R_{ee}$ in the whole system, black from dark to light indicates $\gamma$ from small to large. In (b) and (c), the solid and hollow yellow/green symbols mark $Z$-$R_{ee}$ in the fast/slow band before and after the appearance of shear banding. The color of hollow symbols from dark to light corresponds $\gamma$ increase.

In the linear $Z$-$R_{ee}$ region under all shear conditions,

$$R_{ee} = R_{max} + \beta Z\sigma \quad (1)$$

holds, where the intercept $R_{max}$ and the dimensionless slope $\beta$ can be obtained by linear fitting. The data points used for fitting at different $W_i$ are given in Table 1 of Supporting Information. $R_{max}$ is the end-to-end distance of molecular chain when entanglements are completely released along $Z$-$R_{ee}$ curve in Figure 6. $R_{max}$ decreases linearly with the increase of $ln\,W_i$ as indicated in Figure S9 of supporting information, namely $R_{max} = 310.3\sigma - ln(W_i^{16.8})\sigma$. $\beta$ for different $W_i$ are plotted in Figure 7. $W_i = 5$ just entering the nonlinear rheology region is excluded. $\beta$ attenuates non-linearly with the increase of $W_i$, satisfying

$$\beta = ln(W_i^{0.87}) - \xi_0 \tag{2}$$

The fitting yields $\xi_0 = 11.2$, which is intrigued that $\xi_0$ approaches the initial average entanglement length. For the coupling relationship between chain stretching and disentanglement upon $W_i = 25 \sim 1000$,

$$R_{ee} = R_{max} + \left[ln(W_i^{0.87}) - \xi_0\right]Z\sigma \tag{3}$$

is obtained, which will be discussed later.

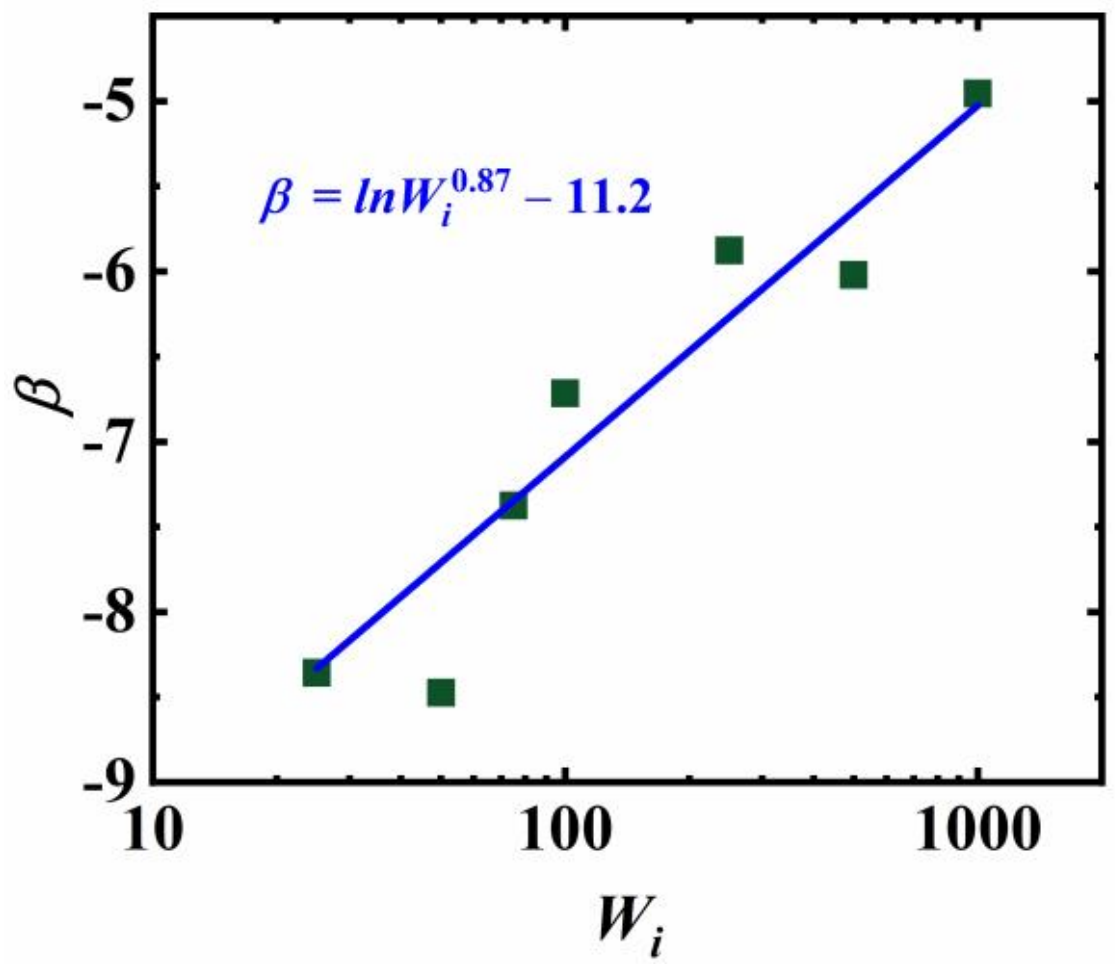

Figure 7 The negative slope of $R_{ee}$ with respect to $Z$ as a function of $W_i$.

**3.3. The thermodynamics in shear.** The free energy changing with shear is calculated in the framework of quasi-equilibrium thermodynamics. The mean free energy $F$ per chain includes the contributions of conformational entropy $S_{\mathrm{con}}$ and internal energy $U$, wherein $U$ includes kinetic energy and potential energy ($U_{\mathrm{nonb}} + U_{\mathrm{bond}} + U_{\mathrm{bend}}$), which can be directly output by LAMMPS software. $S_{\mathrm{con}}$ is given by $-\frac{3}{2}k_{\mathrm{B}}(\frac{R_{\mathrm{ee}}}{\sqrt{N}l})^{2}$, where $N$ and $l$ are the chain length and the bond length. The contribution of entanglement to free energy is not calculated due to the lack of an accepted analytical formula. The variations of $F$ during shear is denoted as $\Delta F$. $\Delta F$ as the function of $Z$ at $W_i$ = 50, 100, 500 and 1000 is plotted in Figure 8, wherein shear causes $Z$ to decrease from its initial maximum and then increase. There is a decrease in internal energy caused by intermolecular slippage during shear startup (see Figure S11 of supporting information for internal energy data). In the early stage of shear, chain stretching cannot immediately offset the change of internal energy, leading to a reduction of $\Delta F$. In later stage of shear, chain stretching dominates and consequently results in the increase of $\Delta F$. When $Z$ reaches the minimum, $\Delta F$ reaches the maximum, where shear banding occurs. Thereafter, $Z$ rebounds with chain retraction, resulting in $F$ decreasing to a stable value. For $W_i$ = 500/1000, the maximum of the free energy is fuzzy due to the weak chain retraction and entanglement reconstruction in the system.

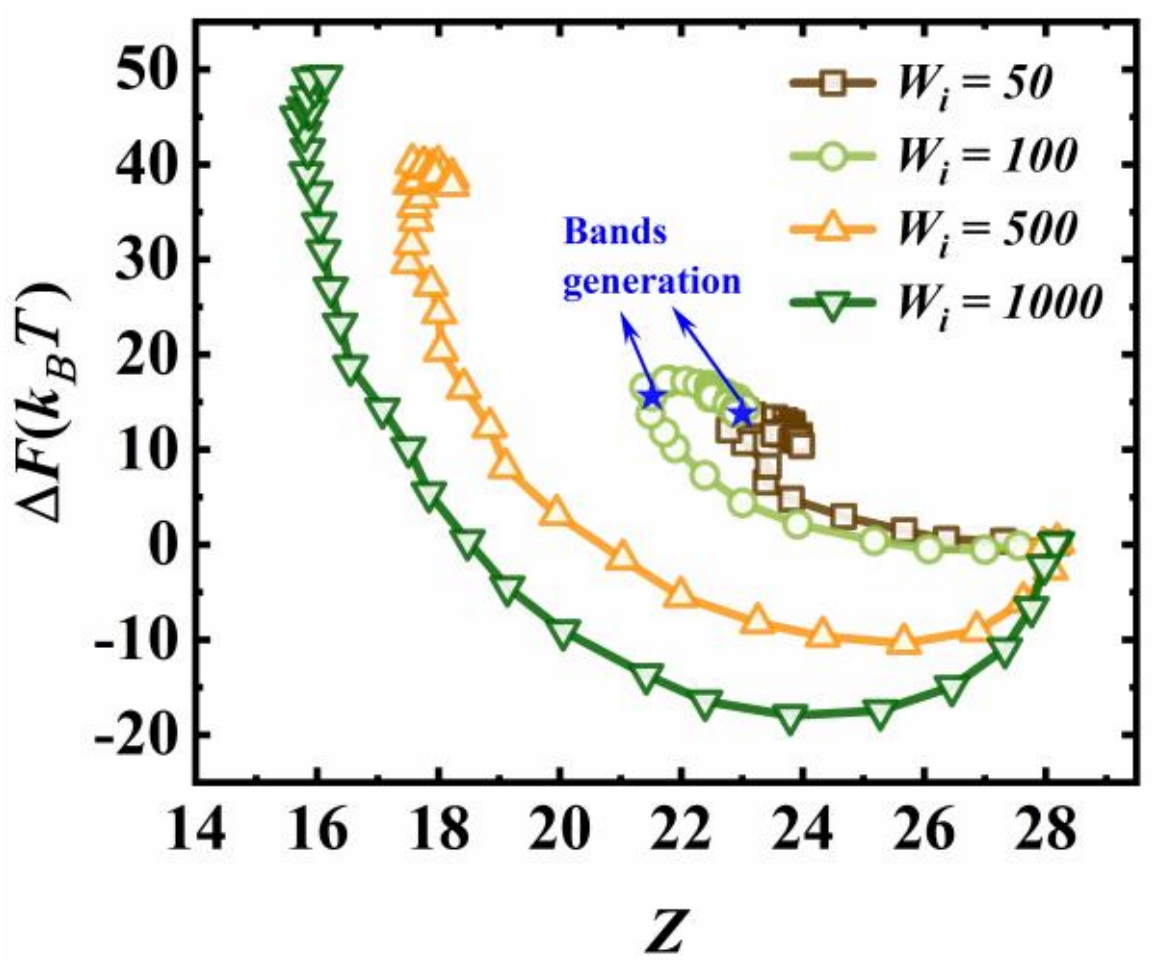

Figure 8 The mean free energy $F$ of a single chain as a function of the average number of entanglement points $Z$ per chain during shearing. The blue star symbols mark the onset of shear banding at $W_i$ = 50/100.

Shear banding is a state with lower free energy than homogeneous flow under shear. In a given shear flow, shear banding is more stable so there is a transition from homogeneous shear to shear banding. In turn, changing the shear flow may make homogeneous shear more stable.[14, 22] To explore whether a transition from shear banding to homogeneous flow can occur, we increase $W_i$ in a system with shear banding already generated at smaller $W_i$. For the system sheared with $W_i$ = 100 and $\gamma$ = 25, we gradually increase $W_i$ from 101 to 1000 with $W_i$ increment of 1 per 0.05 strain. Then the system is continuously sheared at $W_i$ = 1000. Figure 9a shows the changes of shear stress $\sigma_{xy}$ during the whole shearing process. $\sigma_{xy}$ remains on the plateau before $\gamma$ = 25, then increases slightly during $\gamma$ = 25 ~ 70 with increasing $W_i$, and keeps constant in $\gamma$ = 70 ~ 100 with constant $W_i$ = 1000. Figure 9b gives the evolutions of the average end-to-end distance $\langle R_{ee} \rangle$ of the system and the average entanglement number $Z$ per chain during the whole shearing. Increasing $W_i$, $Z$ decreases, while $\langle R_{ee} \rangle$ increases and remains constant around 95$\sigma$ (with $Z$ = 19) at $W_i$ = 1000, which is lower than stable $\langle R_{ee} \rangle$ = 112$\sigma$ (with $Z$ = 17) when $W_i$ = 1000 is directly applied to the static state (see Figure 6d). The velocity profiles at the end of each shear stage are shown in Figure 9c. There seems already having a tendency for four bands to appear at $\gamma$ = 25 with $W_i$ = 100. The four bands are completely formed during the increase of $W_i$ and maintain upon constant $W_i$ = 1000. Figure 9d provides the distribution of $N_Z$ in different layers at the end of each shear stage. The difference between the peaks and valleys of $N_z$ distribution further increases during the shear of increasing $W_i$. At $\gamma$ = 100, the two minima of the $N_z$ distribution are the same and the two maxima are different, which results in two slow bands with different widths. The $Z$-$R_{ee}$ curve during the whole shearing can be seen in Figure S12 of supporting information, wherein the original linear evolution is maintained during the shear of changing $W_i$, which gives a different slope as compared to that sheared directly with $W_i$ = 1000. The simulation of changing shear rate gradually from $W_i$ = 1000 to $W_i$ = 100 is given in Figure S13 of supporting information. Entanglement density gradually increases and redistributes as $W_i$ decreases, but the velocity profile varies weakly with the decrease of $W_i$, which is associated with strong disentanglement at $W_i$ = 1000 in the shearing history.

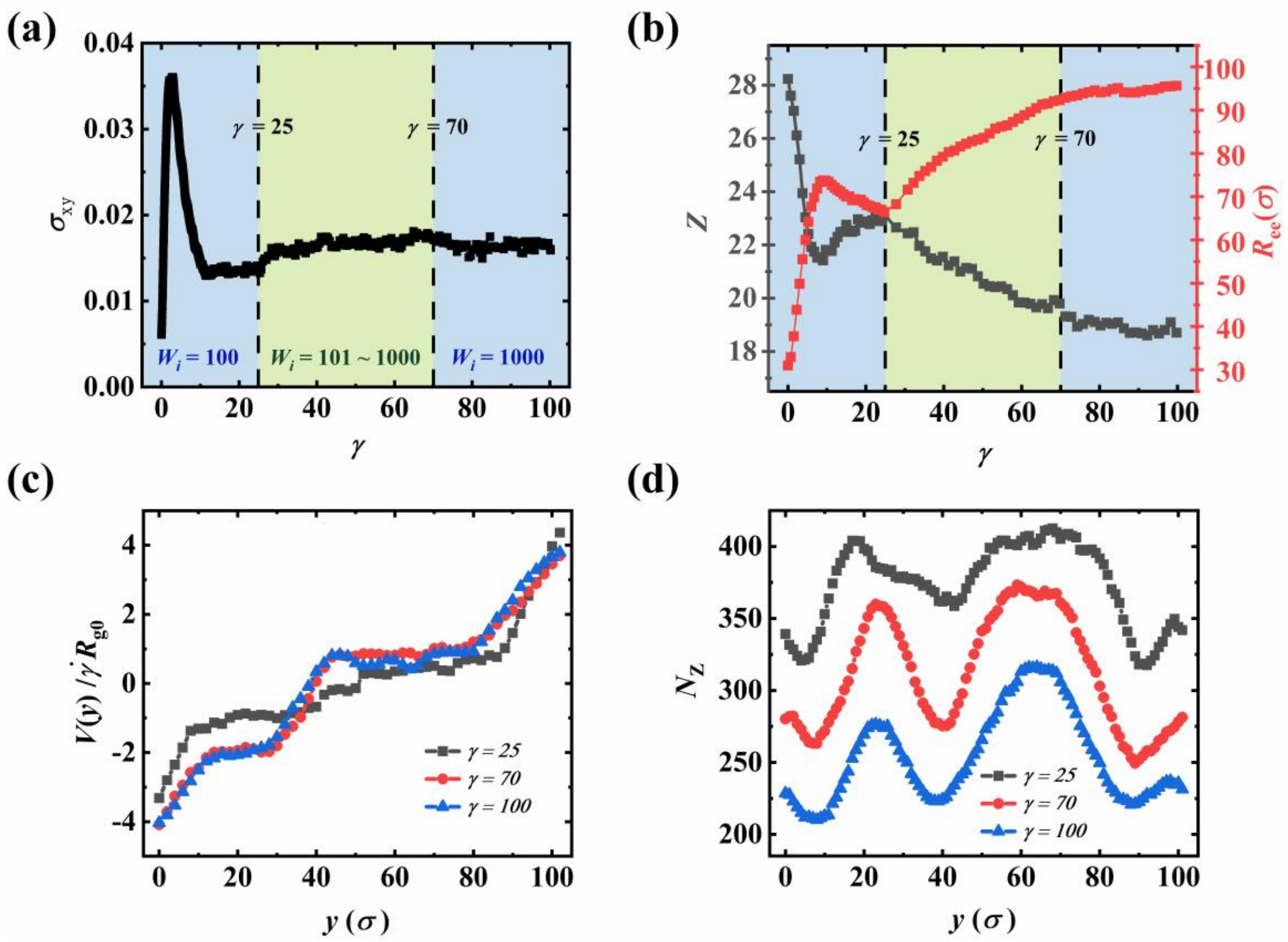

Figure 9 (a) The shear stress as the functions of $\gamma$. In the blue-filled areas, the system is sheared at constant $W_i$ = 100 and 1000. In the red-filled area of $\gamma$ from 25 to 70, $W_i$ is gradually increasing from 101 to 1000. (b) The number of entanglement points per chain (gray symbols) and average end-to-end distance of chains (red symbols) as the functions of $\gamma$. (c) The velocity profile at $\gamma$ = 25, 70, and 100. (d) The number of entanglement points in different layers at $\gamma$ = 25, 70, and 100.

## 4. Discussion

The appearance of shear banding is accompanied by mechanical and thermodynamic instability. Force imbalance on molecular chains as a manifestation of mechanical instability is first considered. There are many kinds of interaction forces for molecular chain in the external field, and some rheological behaviors can be predicted by constructing several forces.[30, 33, 73, 74] Gong et al. give the force-extension curves equation for chains constrained by static topologies, in which the force is proportional to the chain extension and the number of topological charge.[75] Wang et al. proposed the retraction force, entanglement cohesion force, and intermolecular force during molecular deformation under the external flow field,[30, 73] which can explain disentanglement under step deformation and yielding behavior in continuous deformation. Xie et al. proposed a nonequilibrium kinetic theory for entangled polymers, in which blobs in shear flow are subjected to the retraction force and a intermolecular grip force.[33, 74] However, analyzing various detailed forces is a challenge to describe large strain behavior as the configuration changes after shear banding occurs. The subsequent configurational changes after shear banding do not affect the stress as shear banding emerges in the stress plateau region, which makes it more difficult to introduce stress-related molecule forces in detail. Based on the relationship between $Z$ and $R_{ee}$ in Figure 6 and the mean-field idea, the dynamic force balance of the chain during shearing is described by $R_{ee} \sim \left[ln(W_i^{0.87}) - \xi_0\right]Z$.

The effect of entanglement on chain stretching has two sides. On one hand, chain stretching is partly applied by entanglement due to its role as force transmitter, which implies that entanglement promotes chain stretching under flow. On the other hand, it is well known that the maximum stretch ratio of chain is limited by the topological constraint from entanglement, suggesting entanglement hinders chain stretching. Based on the fitting result Eq. (3) obtained in Figure 7, $R_{max}$ is a function of $W_i$. $\left[ln(W_i^{0.87}) - \xi_0\right]Z$ represents the superposition of entanglement effects. The negative value of $\left[ln(W_i^{0.87}) - \xi_0\right]$ indicates the total entanglement effect is to prevent the chain stretching, which corresponds to the inverse relationship between $R_{ee}$ and $Z$ shown in

Figure 7. Therefore, $R_{ee}$ is $R_{max}$ minus the portion of the chain that should be stretched but constrained by entanglements.

The linear relation between $R_{ee}$ and $Z$ reflects a dynamic balance between chain stretching and disentangling. Due to the effects of flow and dynamic entanglement constraint, the increase of $R_{ee}$ involves a variety of molecular forces. In flow, the elastic retract force $f_{retract}$ against deformation is first considered, which is proportional to end-to-end distance $R_{ee}$ [76], namely

$$f_{retract} \sim R_{ee}\frac{K_B T}{\sigma^2} \tag{4}$$

The chain under affine deformation is shown in Figure 10a, where entanglement is maintained on the chain and $f_{retract}$ takes effect. Figure 10b gives the schematic diagram of the affine deformed chain releasing an entanglement, chain stretching or the increase of $R_{ee}$ is accompanied with entanglement slipping and disentanglement. Multiplying the both sides of Eq. (3) with $\frac{K_B T}{\sigma^2}$ and moving the $Z$-related term to the left side, we can get:

$$R_{ee}\frac{K_B T}{\sigma^2} - \left[ln({W_i}^{0.87}) - \xi_0\right] Z\sigma\frac{K_B T}{\sigma^2} = R_{max}\frac{K_B T}{\sigma^2} \tag{5}$$

Then the concept of disentanglement force $f_{disent}$ is introduced, which represents the force exerted on the chain due to entanglement slipping. As $\left[ln({W_i}^{0.87}) - \xi_0\right]$ is negative and $R_{max} > R_{ee}$, the $f_{disent}$ is considered to be in the same direction as chain retraction, satisfying

$$f_{disent} \sim -\beta\frac{K_B T}{\sigma} = -\left[ln({W_i}^{0.87}) - \xi_0\right]\frac{K_B T}{\sigma} \tag{6}$$

Thus each molecular chain suffers $Z$ disentanglement forces $f_{disent}$ imposed on the $Z$ entanglement points and an elastic retract force $f_{retract}$ on the two ends. In the homogeneous flow and shear banding, Eq. (5) holds. Under a given $W_i$, $R_{max}$ is constant, then a dynamic force balance holds during flow, namely

$$f_{retract} + Zf_{disent} = constant \sim R_{max}\frac{K_B T}{\sigma^2} \quad (7)$$

, where the *constant* term is equivalent to the retract force when chains are completely untangled ($Z = 0$). As different $W_i$ correspond to different $R_{max}$, $R_{max}$ represents the response property of molecular conformation under flow, during which disentangling and stretching of chain couples together. Under shear with higher $W_i$, the dynamic force balance holds at a lower force level as $R_{max}$ decreases with the increase of $W_i$. The relation indicated by Eq. (7) suggests a dynamic compensation between $f_{retract}$ and $Zf_{disent}$ takes place during disentanglement process in the homogeneous and shear banding flow. In another word, the reduction of the total disentanglement force $Zf_{disent}$ due to releasing entanglement (the decrease of $Z$) is precisely taken over by $f_{retract}$ in this process. Otherwise, force balance is broken, resulting in the enhancement of structural heterogeneity. This happens in the transition flow region, where the molecular chain undergoes a force imbalance and $Z$-$R_{ee}$ deviates from the linearity as shown in Figure 6.

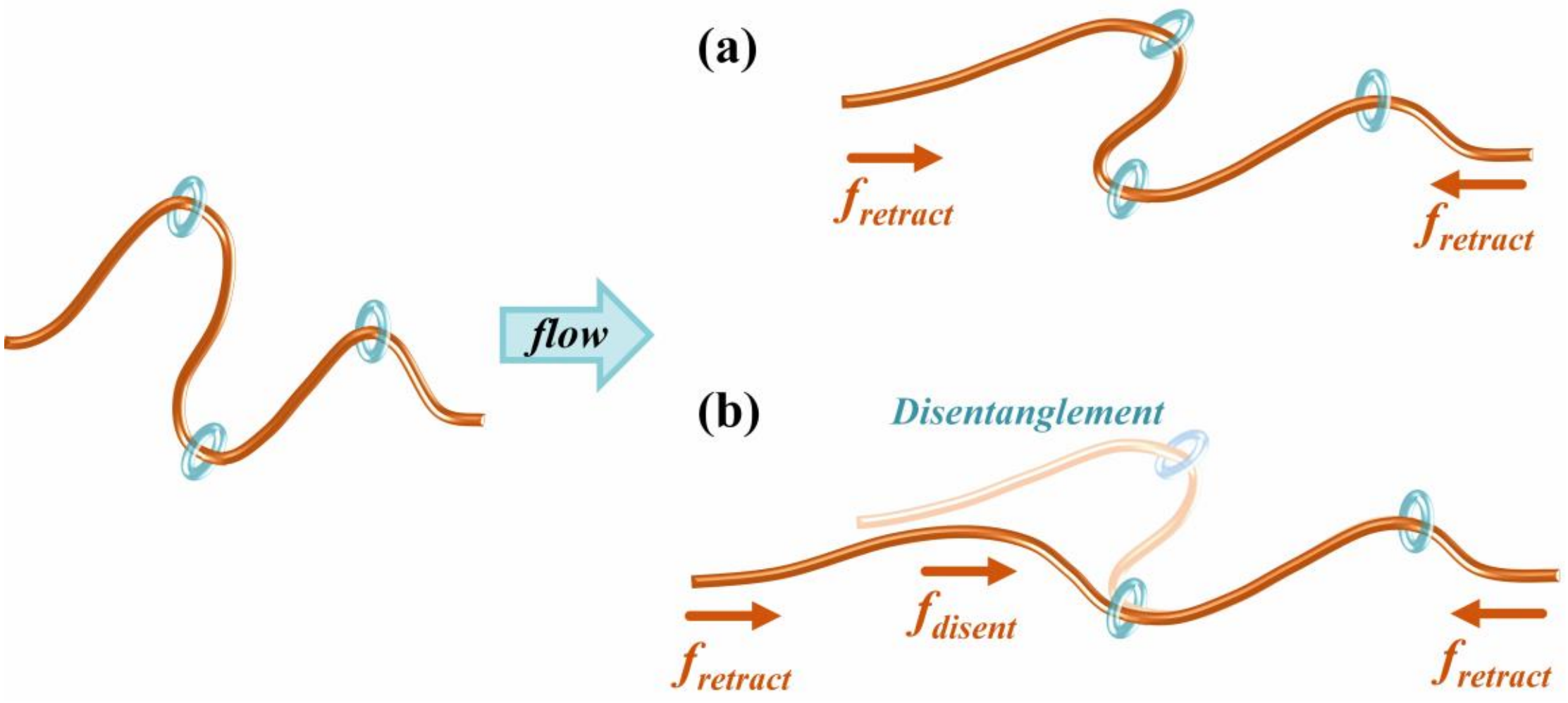

Figure 10 The molecular chain in the initial and flow state. (a) Affine deformation of chain. (b) The chain releases an entanglement during affine deformation. The blue circles represent entanglement constraints.

From the perspective of thermodynamic, disentanglement requires activation of entanglement to slip certain distance, which should be in the order of monomer scale $\sigma$. As disentanglement force given by Eq. (6) and $-\beta\sigma$ represents the increase in $R_{ee}$ when an entanglement is released under flow, suggesting that $-\beta$ can be regarded as the dimensionless activation energy. Then $\beta = \left[ln(W_i^{\,0.87}) - \xi_0\right]$ can be rewritten as

$$\exp(\beta) \sim W_i^{\,0.87} \tag{8}$$

Here we introduce the activation energy $E_d$ of disentanglement driven by shear flow, the disentanglement rate can be given by

$$V_d \sim \exp(-E_d/k_BT) \tag{9}$$

Note that $V_d$ is not the rate with time, but the rate with chain stretching. Considering $E_d$ is achieved with chain being stretched for $-\beta\sigma$, there is

$$-\beta \sim E_d/k_BT \tag{10}$$

Combining Eq. (8), we have

$$V_d \sim W_i^{\,0.87} \tag{11}$$

for homogeneous flow and shear banding under high shear rate.

The calculation method of entanglement affects the numerical result of entanglement, but does not affect the scaling relationship between $\beta$ and $W_i$. Repeating all the entanglement calculations with Z1,[67-70] the results are not fundamentally different from those obtained with Z1+, but with an overall decrease of about 2.2. The $Z$-$R_{ee}$ curves under different $W_i$ and the fitting of $\beta$ and $R_{max}$ based on Z1 are shown in Figure S10 of Supporting Information, which also gives $\beta = ln(W_i^{0.87}) - \xi_0$ with $\xi_0$ = 11.1.

The generation of shear banding can also be considered as a quasi-phase transition to reduce the free energy enhancement caused by shear. The classical phase separation requires the free energy of the system to be reduced, but the phase separation in the system far from equilibration cannot simply be predicted by the reduction of spontaneous free energy alone. The elastic energy of the single-component polymer chain in the flow field is stored and the free energy is increased, which changes the static phase equilibrium of the system.[77] In the monodisperse polymer melt, the mixed free energy is lacking, the conformational entropy elasticity dominates the free energy as Figure 8 shows. The free energy increases before shear banding occurs, and decreases rapidly after shear banding occurs, which is like a first-order phase transition process.

Based on Figure 9, the generation of shear banding endows the system a lower free energy than that of the homogeneous shear. Shear bands are not significant when directly imposing shear at $W_i$ = 1000. For the system with visible bands, the bands with sharp edges remain when increase shearing rate to $W_i$ = 1000 (see Figure 9). This result indicates that shear banding is more stable than homogeneous shear. From the view of thermodynamics, the free energy of shear banding is lower than that of homogeneous shear, which is confirmed by comparing elastic energy storage. $R_{ee}$ in the system with $W_i$ slowly increased to 1000 is smaller than that in the system directly imposed a shear of $W_i$ = 1000, while the former shows more pronounced shear banding than the latter does.

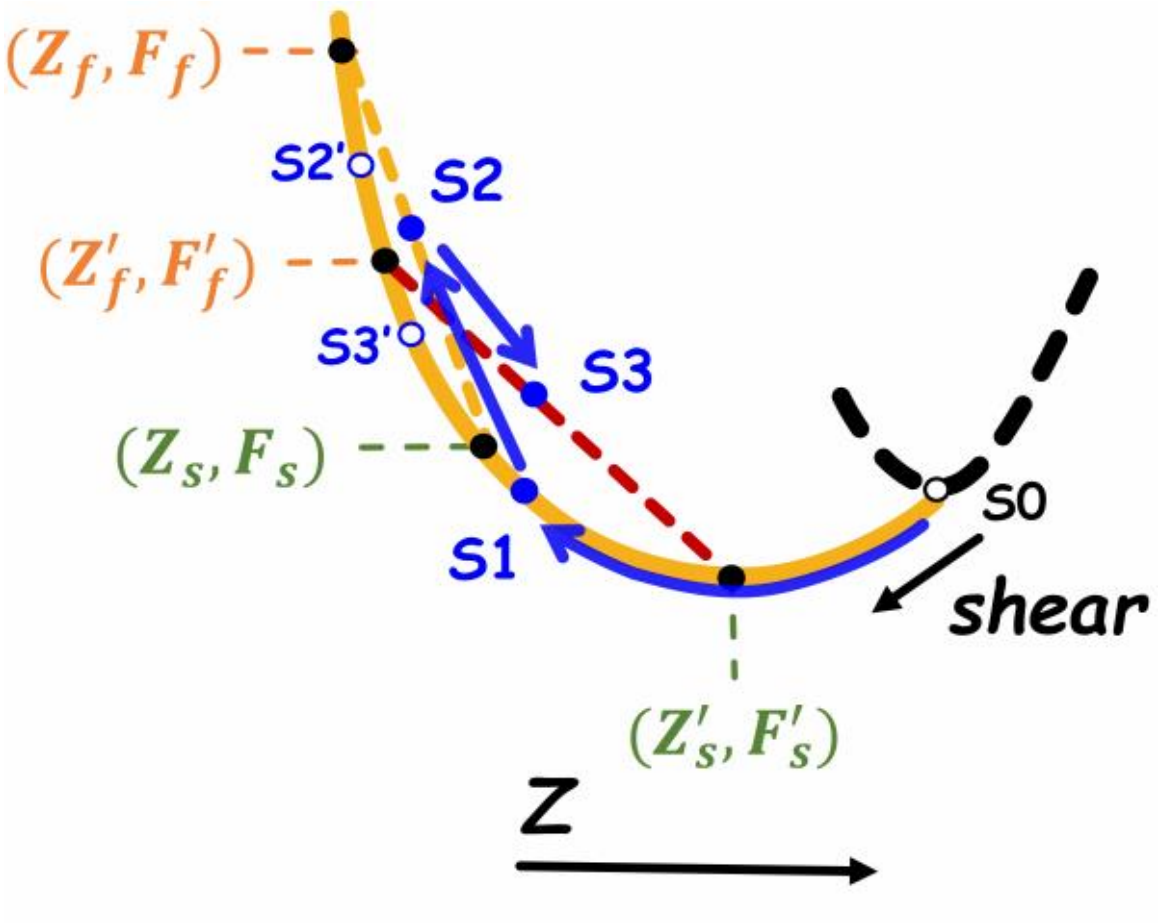

Figure 11 The schematic diagram of the state evolution of molecular chains in shear. The yellow curve represents the free energy landscape of homogeneous shear. $S1$, $S2'$, and $S3'$ represent the states of homogenous shear, $S2$ and $S3$ represent the states during shear banding formation.

Based on the calculated $F$ in Figure 8, a schematic comparison of the free energies between shear banding and homogeneous shear is provided in Figure 11. The black dashed line is the static free energy landscape, $S0$ is the initial system in equilibrium state. The yellow curve shows $F$ changes with $Z$ at homogeneous shear. With chain stretching and disentanglement, $F$ rises to $S1$, where entanglement heterogeneity emerges. In the process from $S1$ to $S2$, the entanglement heterogeneity gradually increases until shear banding occurs at $S2$. Now the system is divided into the fast and slow bands where the corresponding average entanglement are $Z_f$ and $Z_s$, and the state function densities are $F_f$ and $F_s$, i.e. [$Z_f$, $F_f$] and [$Z_s$, $F_s$], respectively. $S2$ is an unstable state, chain retraction and entanglement reconstruction occur to decrease the free energy, during which stable $S3$ is reached, and the system is divided into the fast band of [$Z_f'$, $F_f'$] and the slow band of [$Z_s'$, $F_s'$]. On the $F$ evolution path during shear, the chains are stretched from $S1$ to $S2$, while the chain retracts at $S2'$ and finally reaches stable state $S3'$. From Figure S13b of supporting information and Figure 9, $Z$ at $S2$ and $S3$ are smaller than those at $S2'$ and $S3'$, respectively. The occurrence of shear banding inhibits the upward trend of free energy and dissipates a large amount of free energy, so the system achieves lower free energy by forming shear banding. Nevertheless, $F$ at shear banding state is larger than that of the homogeneous system at quiescent, shear banding cannot occur in the absence of an external field. Instead, the uniformity will be restored when removing the shear field.

The stability of shear bands is associated with bandwidth. The total free energy of the system is equal to $N_{chains}[\varphi_s F_s + (1 - \varphi_s)F_f]$, where $\varphi_s$ is the ratio of the width of the slow band and $N_{chains}$ is the number of chains in the system. Since the free energy of the slow band is lower than that of the fast band, the total free energy of the system becomes lower when $\varphi_s$ increases. At $W_i$ = 50 and 100, the ratio of the width of slow band is constant and larger than that of fast band, a lower free energy is achieved to maintain shear banding. At $W_i$ = 500 and 1000, the small width of slow band makes the system remain in a state with relatively high free energy. The systems are more easily regressed to homogeneous shear under thermal disturbances. Otherwise, there are more bands at higher shear rates, the curve of entanglement distribution tends to be wavy-shaped (see Figures 3e and 3f), which possibly weakens shear banding and reduces the stability of bands compared with well-shaped ones (see Figures 2e and 2f).

The failure to form visible shear banding at high $W_i$ may be due to the higher magnitude of the entanglement distribution fluctuation required, which is prevented by the surface tension resulting from the free energy difference between bands. Considering the heterogeneous free energy distribution, the surface tension between the bands will be generated, which will inhibit shear banding. For the visible shear banding, the stronger chain retraction and entanglement reconstruction inside the bands reduce the total free energy, which competes with surface tension. Additionally, shear rate and bandwidth do not satisfy the lever principle of gas-liquid coexistence,[36, 57] which may be due to shear banding is not a phenomenon strictly dependent on shear rate. The phase diagram of shear banding and homogeneous shear cannot be divided strictly by shear rate and strain only[78], other parameters must be considered such as entanglement concentration fluctuation and shear history.

## 5. Conclusions

In conclusion, the well-entangled polymer melt sheared at different rates were performed by DPD simulation. Shear banding occurs at $W_i$ = 50 ~ 1000. The appearance of shear banding is accompanied by the heterogeneity of entanglement, and the response of velocity to the variation of the entanglement distribution is demonstrated to be instantaneous. The occurrence of shear banding is traced to molecular force imbalance. The criterion for stable shear flow at either homogeneous or shear banding states is observed to be $R_{ee} \sim \left[ln(W_i^{0.87}) - \xi_0\right]Z$, where dynamic force balance is held. Under stable flow, the disentanglement rate is scaling with flow strength as $V_d \sim W_i^{0.87}$.

The formation of shear banding partially dissipates the free energy stored in the system, thus achieving lower free energy than the homogeneous shear. The width of bands will determine the stability of the bands by affecting the free energy of the system. Regarding shear banding as a non-equilibrium phase transition, the difference of free energy between homogeneous shear and shear banding is the driving force of transition, but understanding the nature of shear banding still requires further advances in non-equilibrium phase transition theory.

## ASSOCIATED CONTENT

### Supporting Information

The mean square internal distances of molecular chains in initial system, the shear stress-strain curves at different $W_i$, the entanglement evolutions of short chain systems during shear, the changes of entanglement as strain at $W_i = 5$, the ratio of stress at shear startup stage to the stress overshoot value, the evolutions of distribution of velocity and entanglement at $W_i = 25/75/250$, the intercept of $Z$-$R_{ee}$ curve, the $Z$-$R_{ee}$ curves and fitting results at different $W_i$ based on Z1, the average internal energy of each monomer varies strain, the changes of stress and entanglement during the shear of changing $W_i$.

### Notes

The authors declare no competing financial interest.

## ACKNOWLEDGMENTS

We would like to thank Prof. Yuyuan Lu and Dr. Yongjin Ruan (Changchun Institute of Applied Chemistry, Chinese Academy of Sciences) for their help to realize the simulation of the shear banding and Prof. Martin Kröger (Eidgenössische Technische Hochschule Zürich) for his support with the Z1+ algorithm. This work is supported by the National Key R&D Program of China (2020YFA0405800), the National Natural Science Foundation of China (51890872).

## Supporting Information for

# Shear banding in monodisperse polymer melt

**Fan Peng, Renkuan Cao, Cui Nie, Tingyu Xu * and Liangbin Li***

*National Synchrotron Radiation Laboratory, Anhui Provincial Engineering Laboratory of Advanced Functional Polymer Film,CAS Key Laboratory of Soft Matter Chemistry, University of Science and Technology of China, Hefei 230026, China*

## Including:

## Figure S1-S13

## Table S1, S2

Corresponding authors, E-mail: tyxu@mail.ustc.edu.cn (T. Y. Xu); lbli@ustc.edu.cn (L. B. Li)

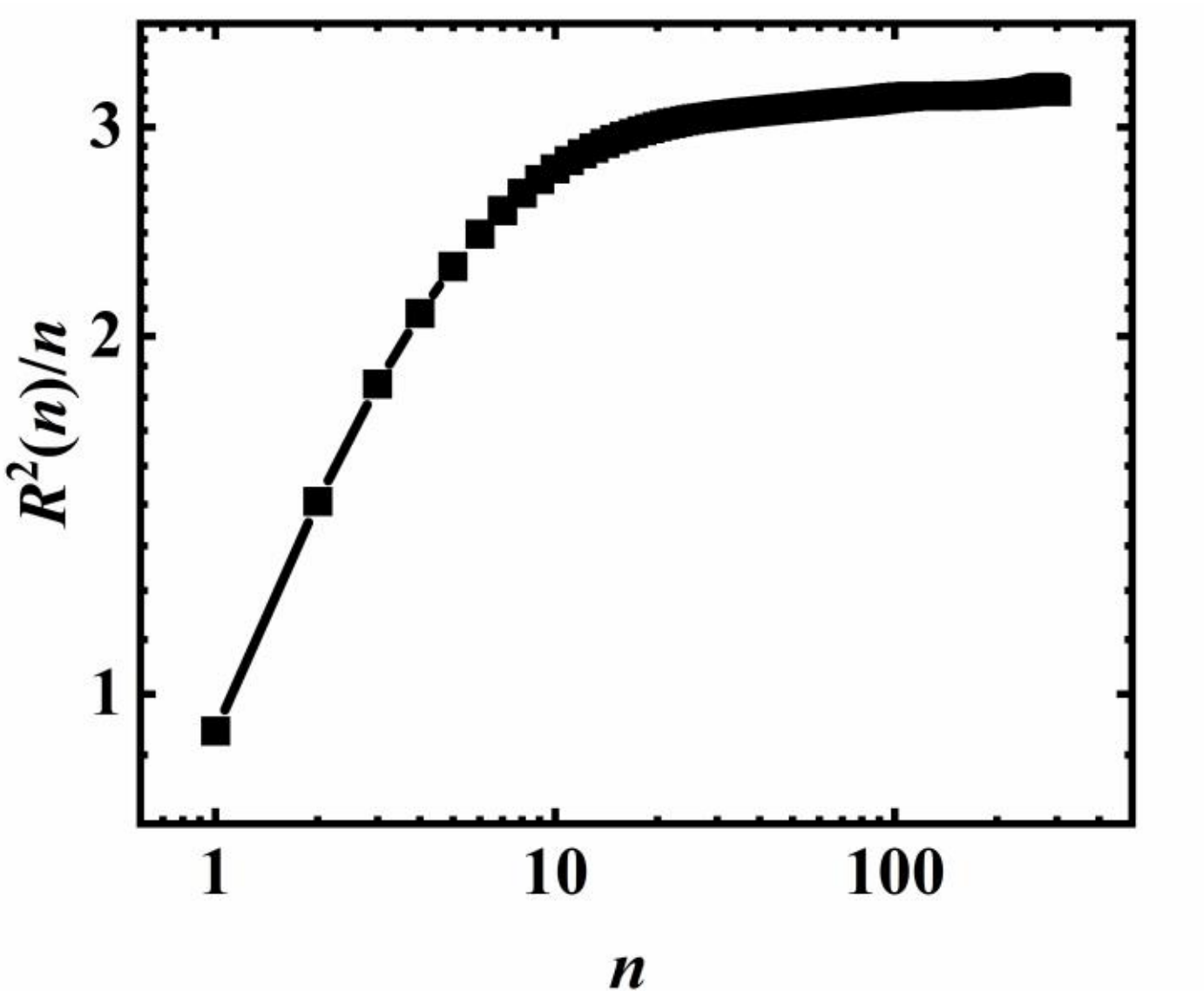

Fig. S1. The mean square internal distances of molecular chains in initial system.

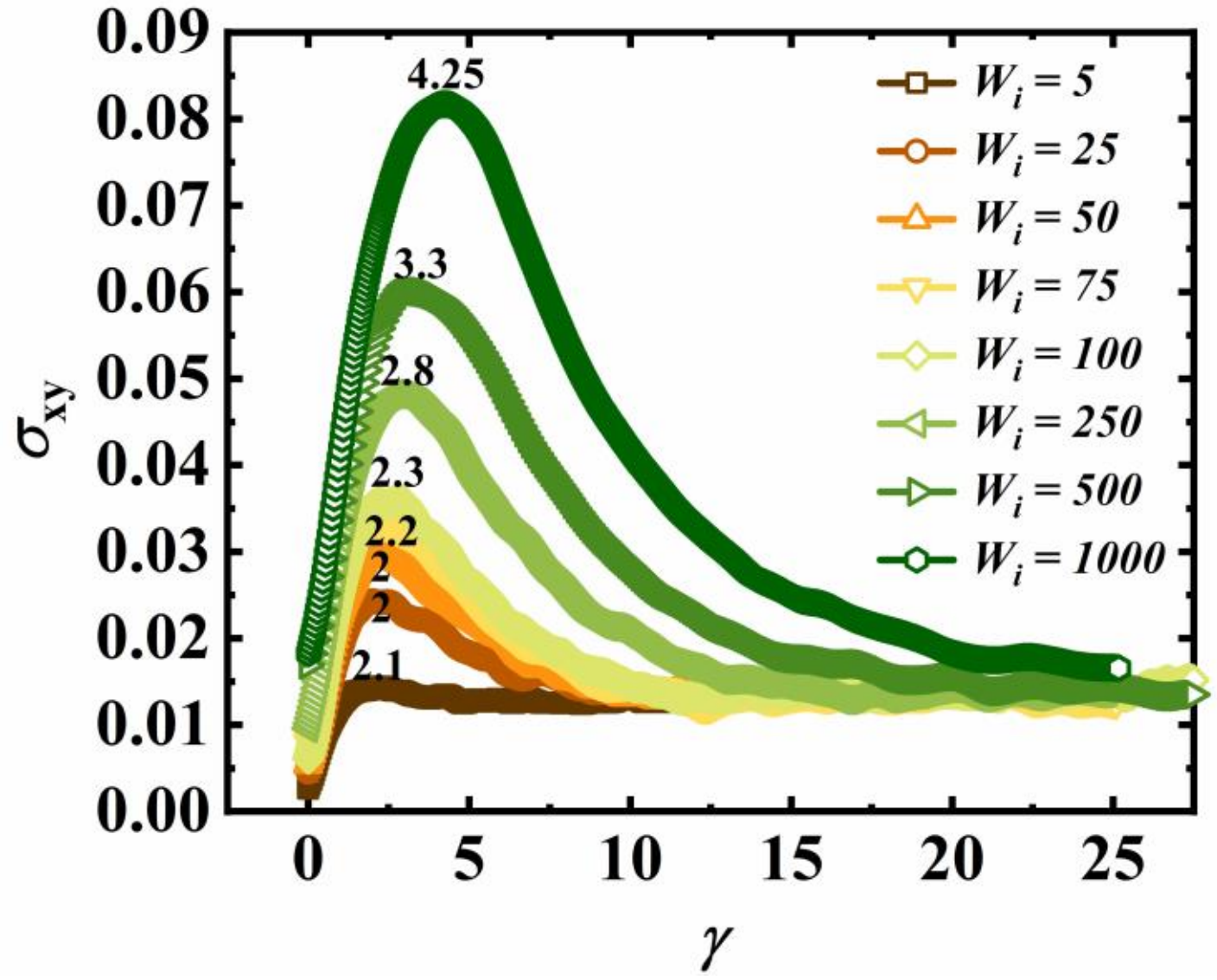

Fig. S2. The shear stress $\sigma_{xy}$ as the function of strain $\gamma$ at different $W_i$. The black numbers in the figure give the strain at the maximum stress.

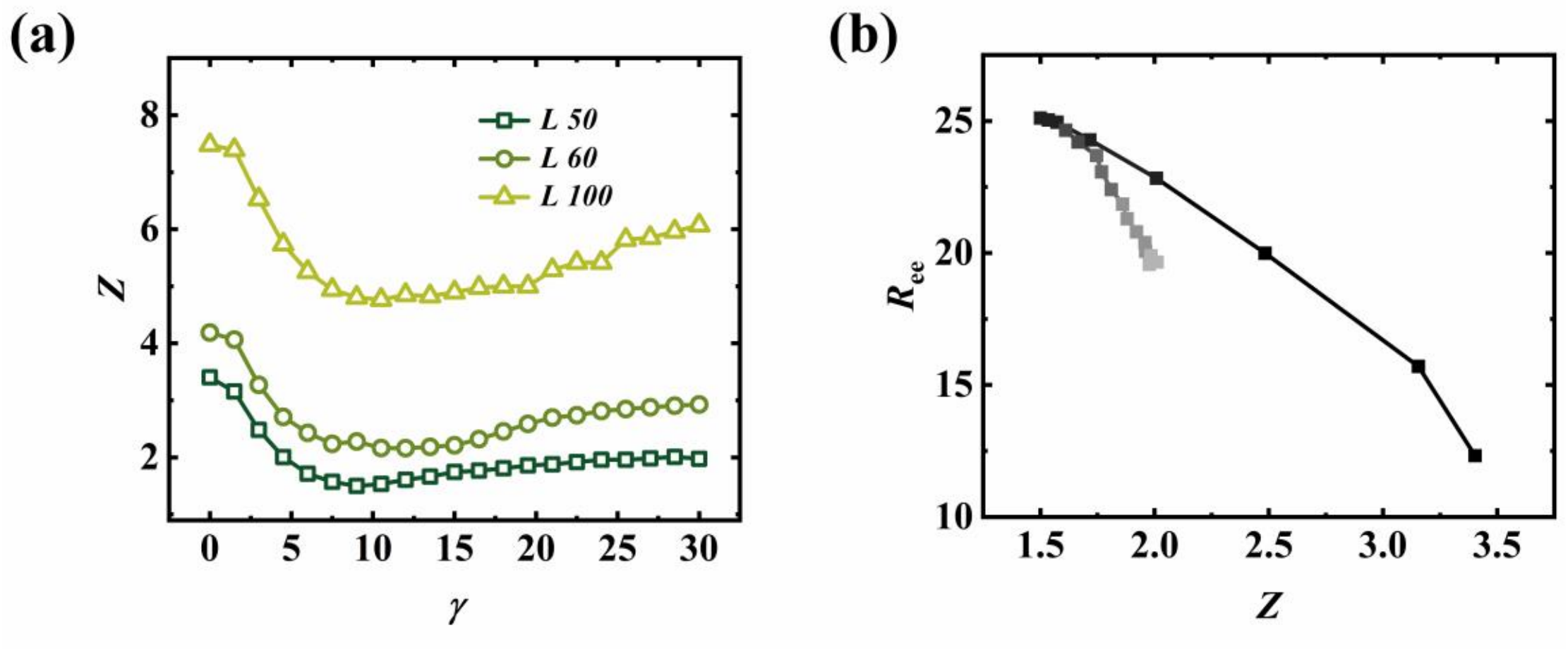

Fig. S3. The system with chain length of 50/60/100 is sheared at $W_i = 50$. The simulation conditions during shear are the same as which described in **Simulation Method**. The number of monomers in the whole system is constant 493800. (a) The average number of entanglements $Z$ of a single chain varies with strain. (b) The $Z$-$R_{ee}$ curve in the system with chain length of 50. The color of hollow symbols from dark to light corresponds $\gamma$ increase.

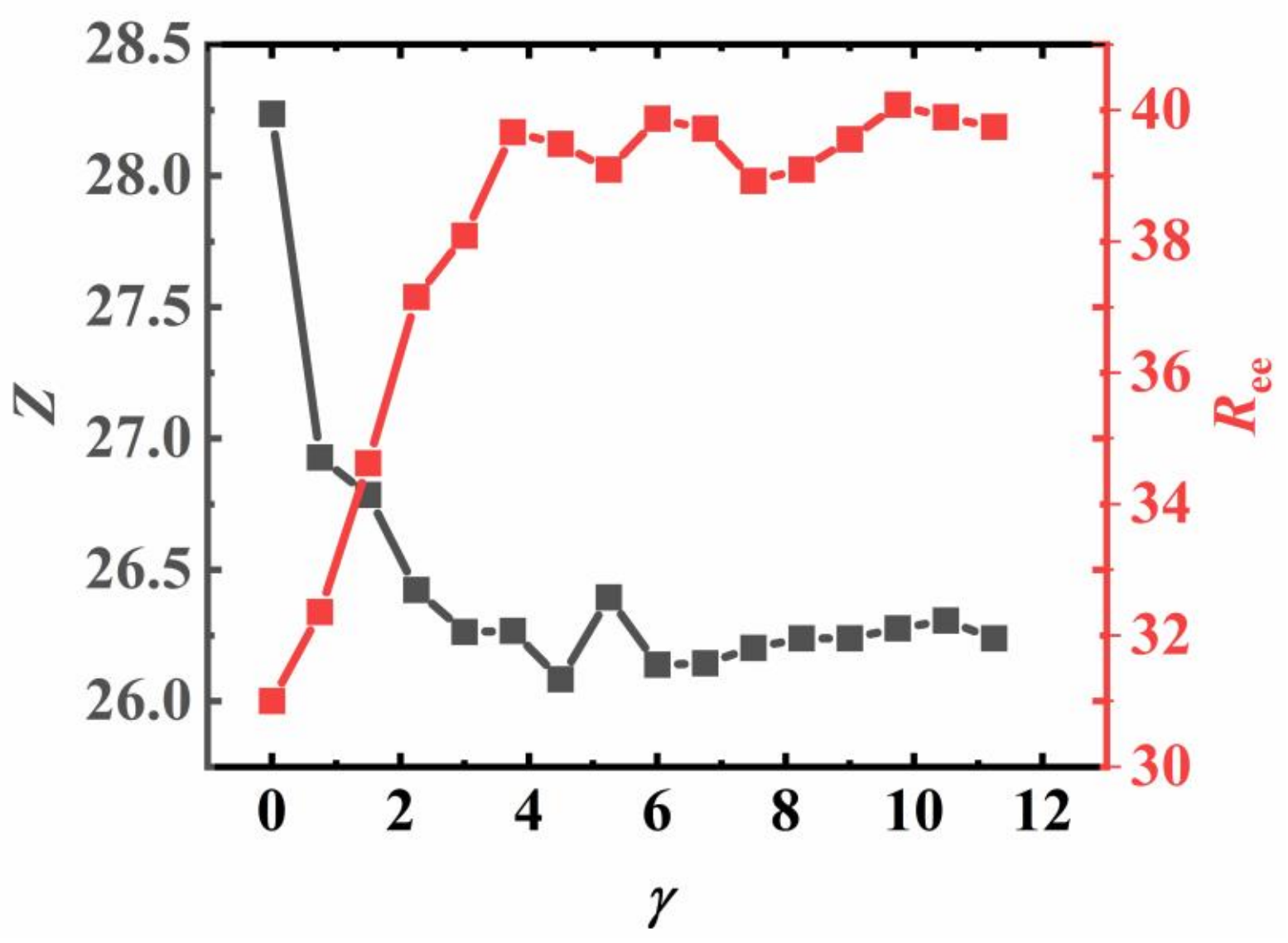

Fig. S4. The $R_{ee}$ and $Z$ as the function of strain during shear of $W_i = 5$.

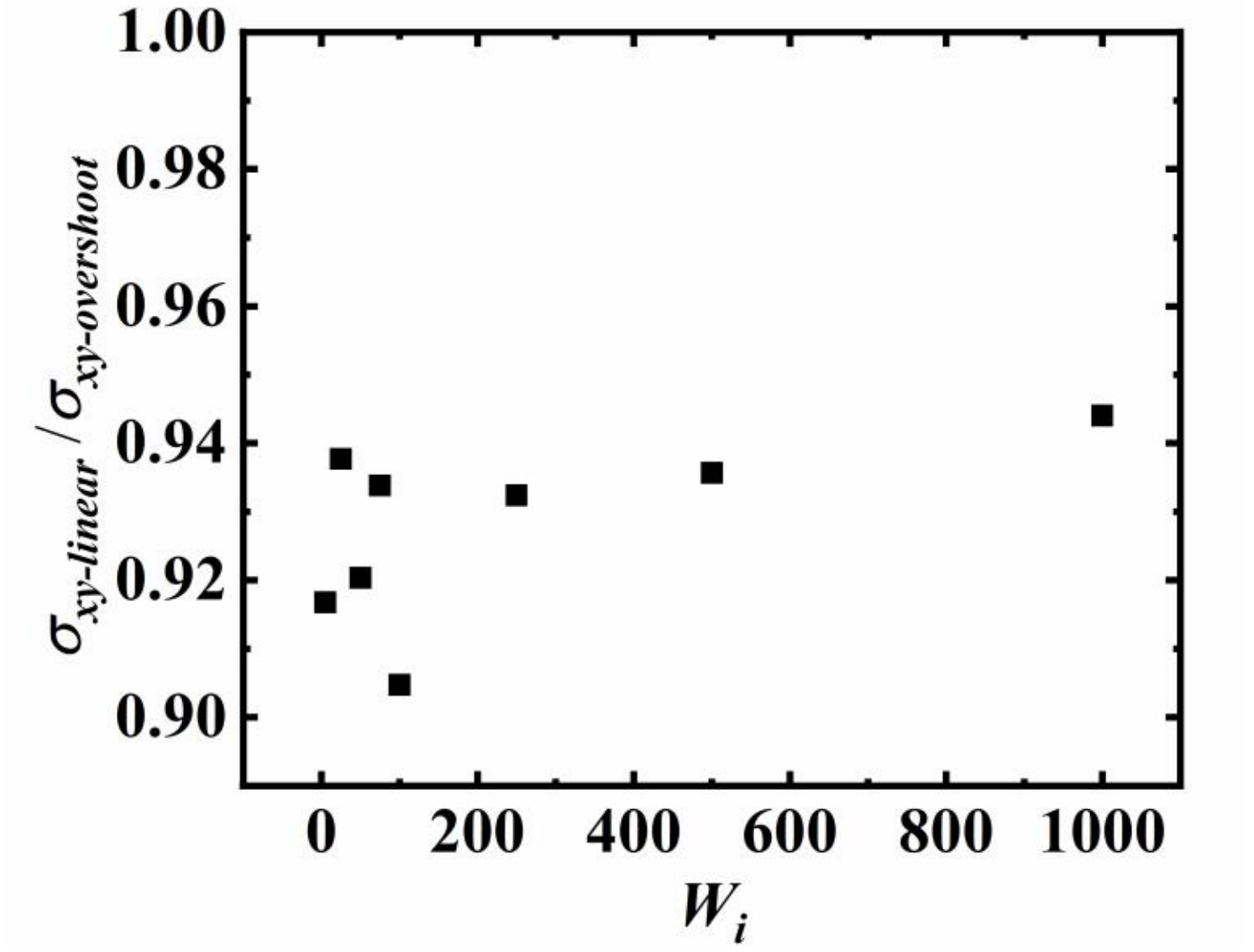

Fig. S5. The ratio of the stress when $Z$-$R_{ee}$ first enters the linear region ($\sigma_{xy\text{-linear}}$) to the stress when overshoot occurs ($\sigma_{xy\text{-overshoot}}$) at different $W_i$, where the corresponding strain and stress are shown in Table S1.

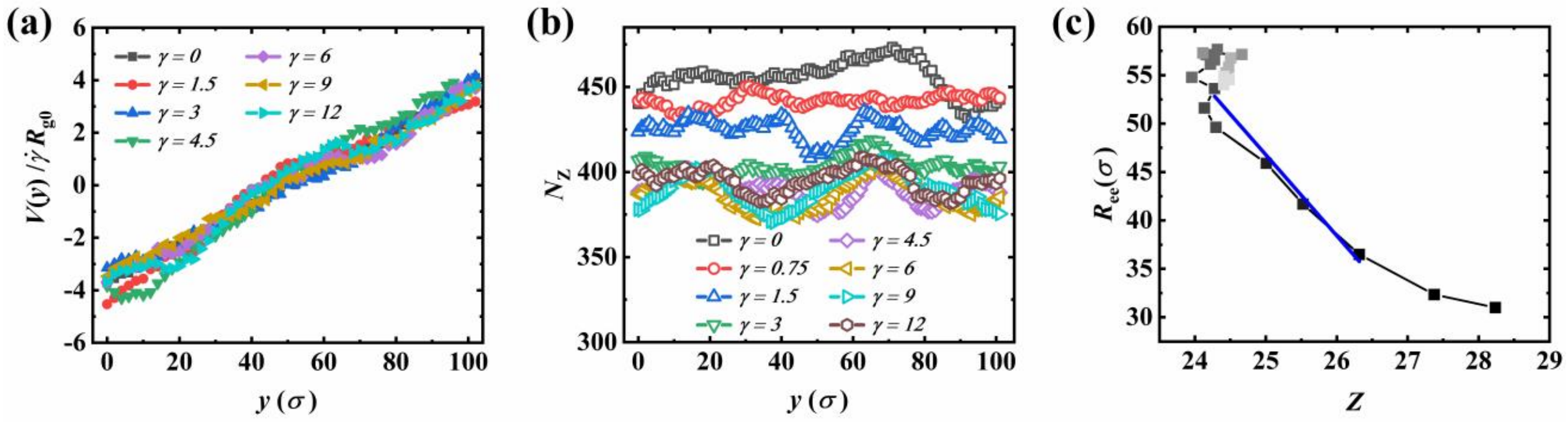

Fig. S6. The (a) velocity profile, (b) distribution of entanglement points on the direction of velocity gradient, (c) $Z$-$R_{ee}$ curve during shear of $W_i$ = 25.

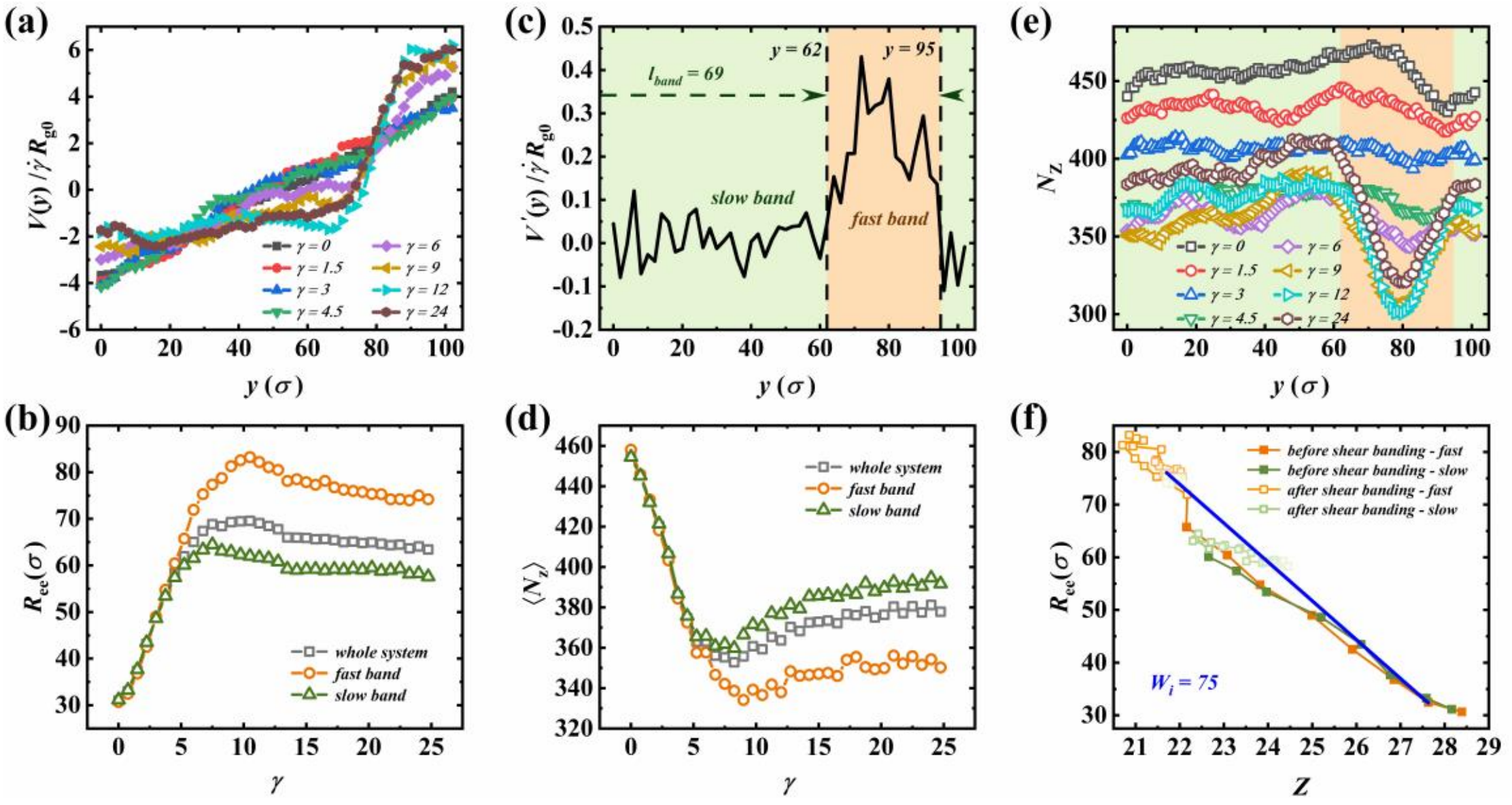

Fig. S7. The (a) velocity profile, (b) distribution of entanglement points on the direction of velocity gradient, (c) $Z$-$R_{ee}$ curve during shear of $W_i$ = 75.

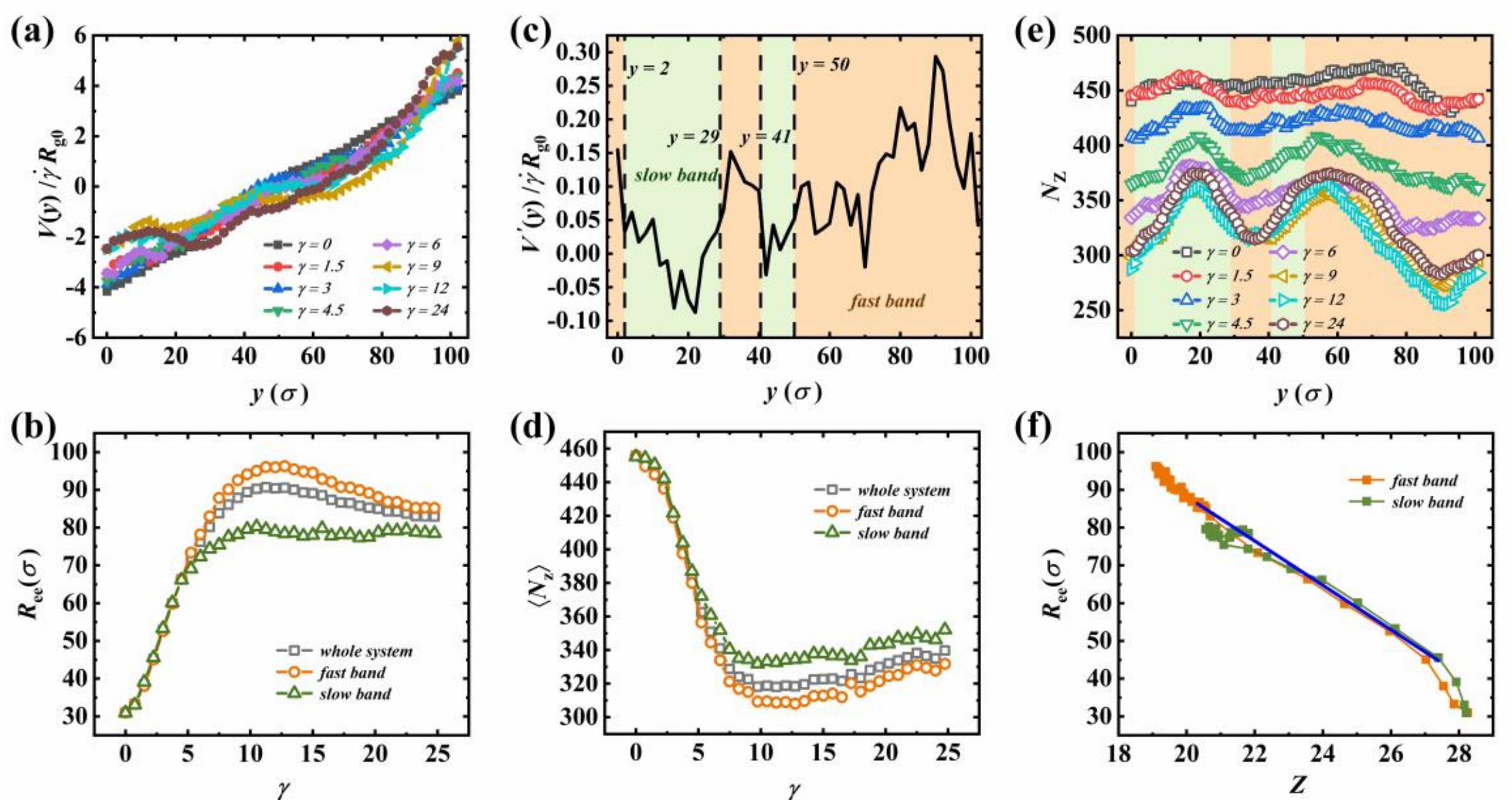

Fig. S8. The (a) velocity profile, (b) distribution of entanglement points on the direction of velocity gradient, (c) $Z$-$R_{ee}$ curve during shear of $W_i$ = 250.

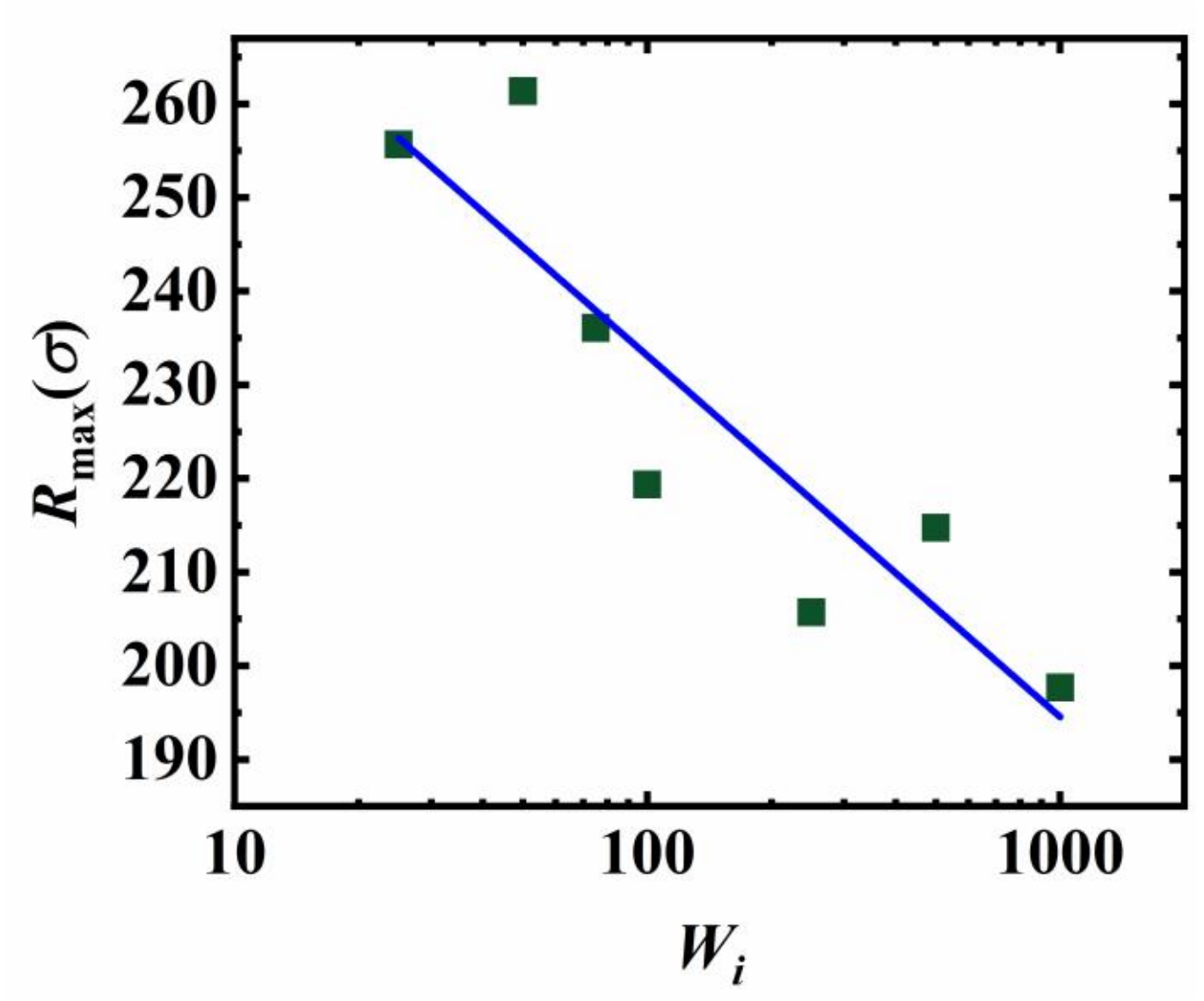

Fig. S9. The intercept of $Z$-$R_{ee}$ curve as the function of $W_i$, the relationship satisfies $R_{max} = 310.3\sigma - ln(W_i^{16.8})\sigma$.

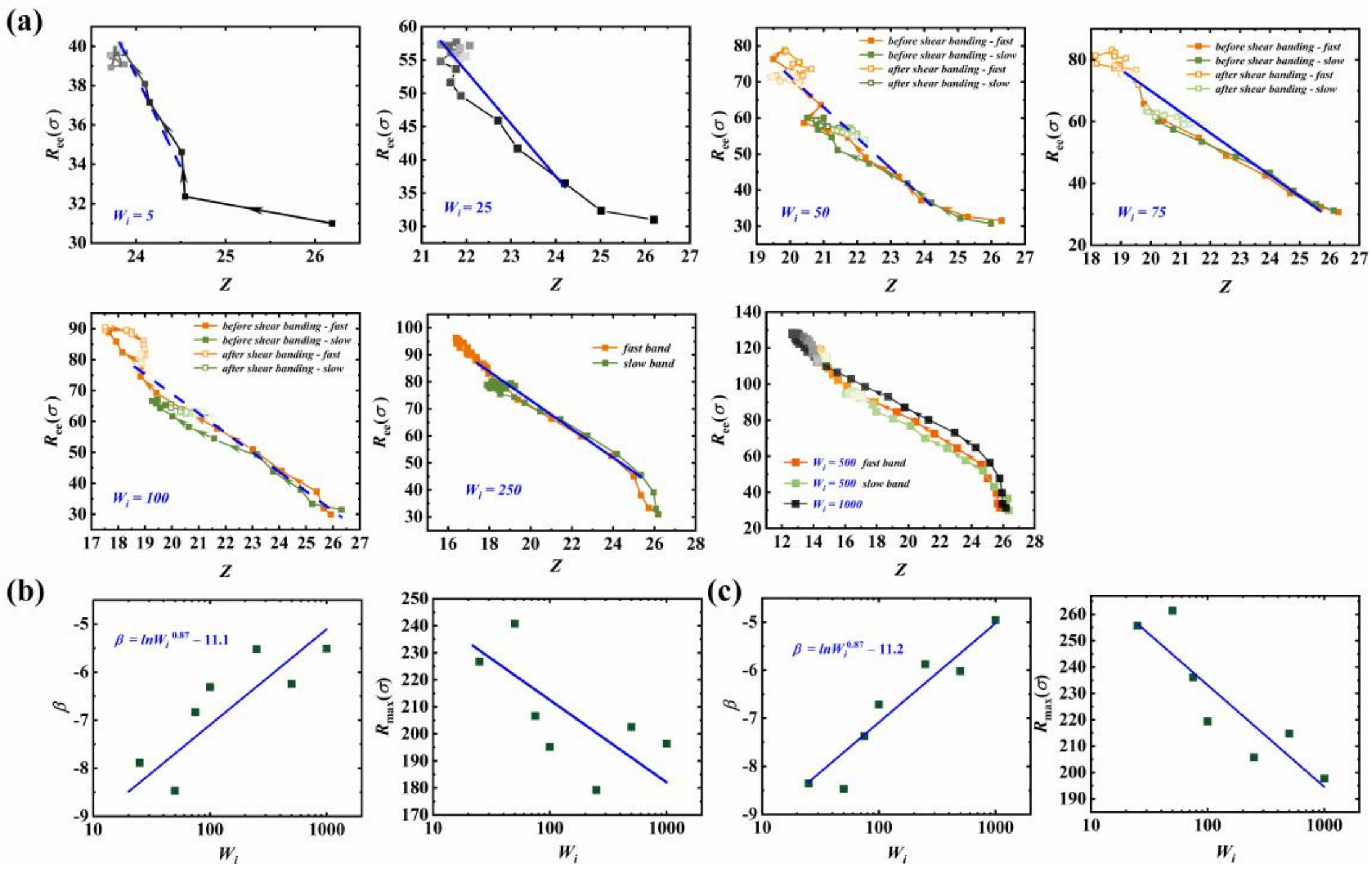

Fig. S10. (a) The $Z$-$R_{ee}$ curve in different $W_i$, where the entanglement data is calculated using Z1. (b) Based on Z1, the slope $\beta$ and intercept $R_{max}$ of $Z$-$R_{ee}$ curve as the function of $W_i$. (c) Based on Z1+, the fitted $\beta$ and $R_{max}$ change as $W_i$.

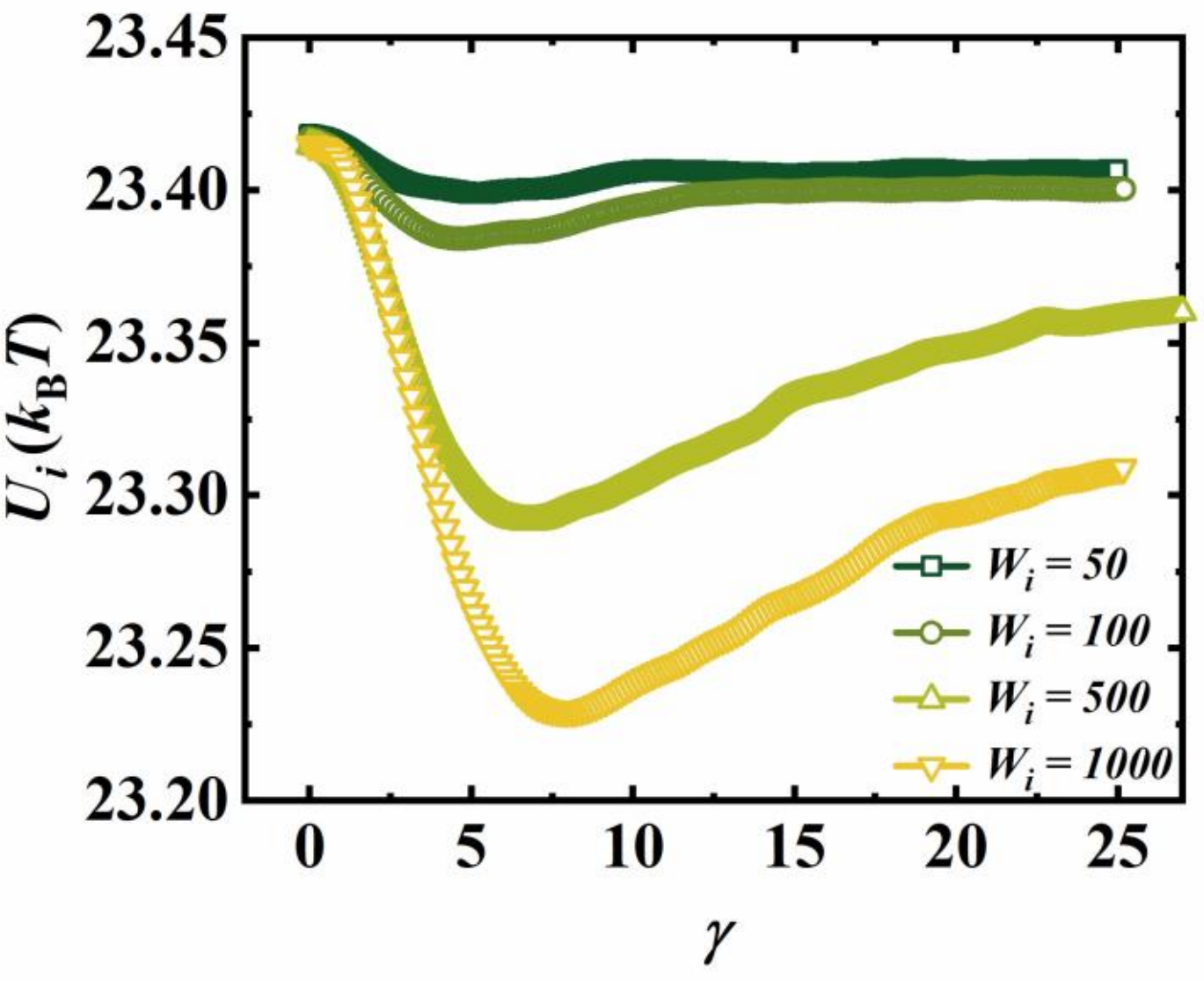

Fig. S11. The average internal energy of each monomer varies strain.

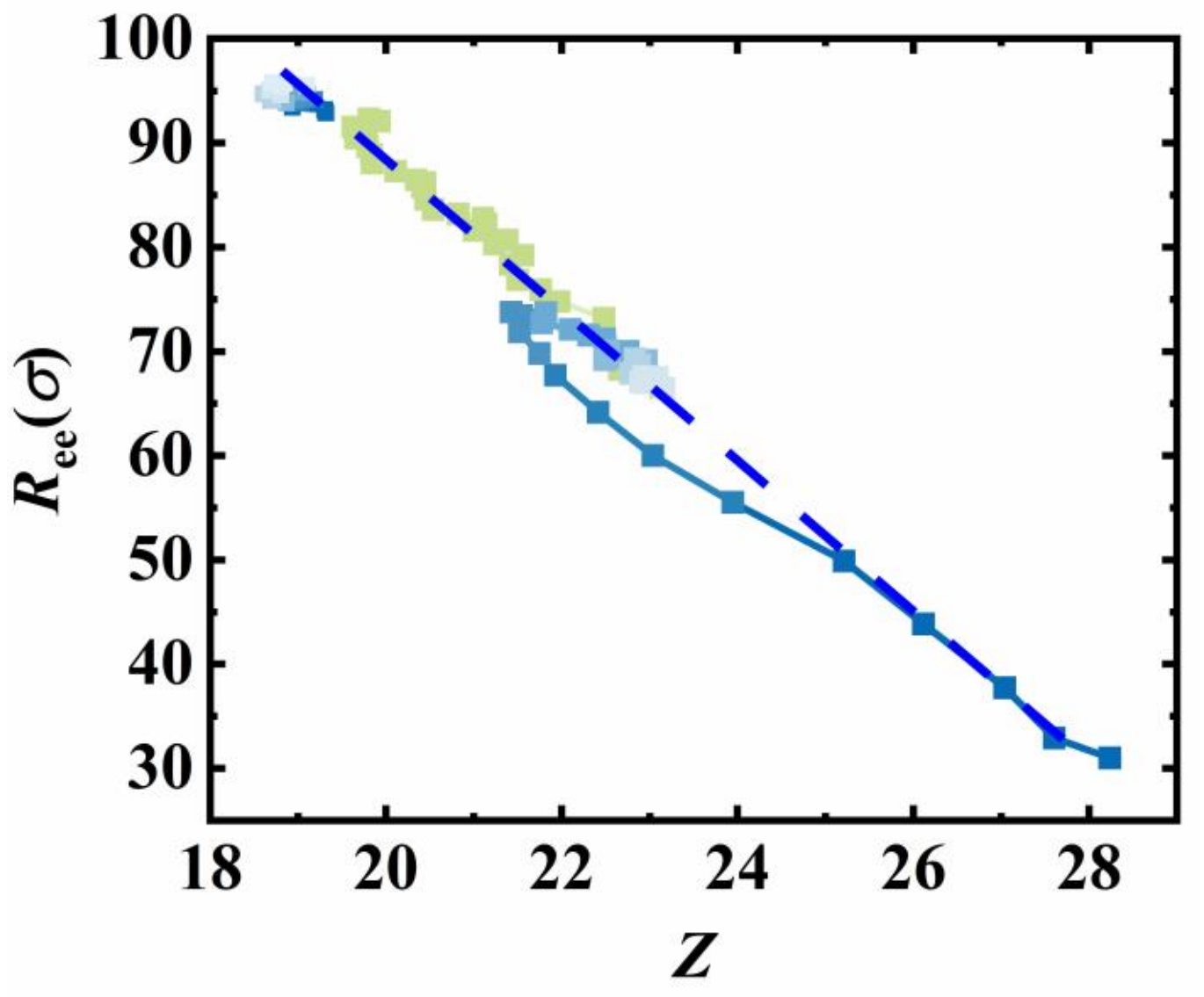

Fig. S12. $R_{ee}$ as the function of $Z$ during shear of $W_i$ =100, $W_i$ =101 ~ 1000 and $W_i$ =1000.

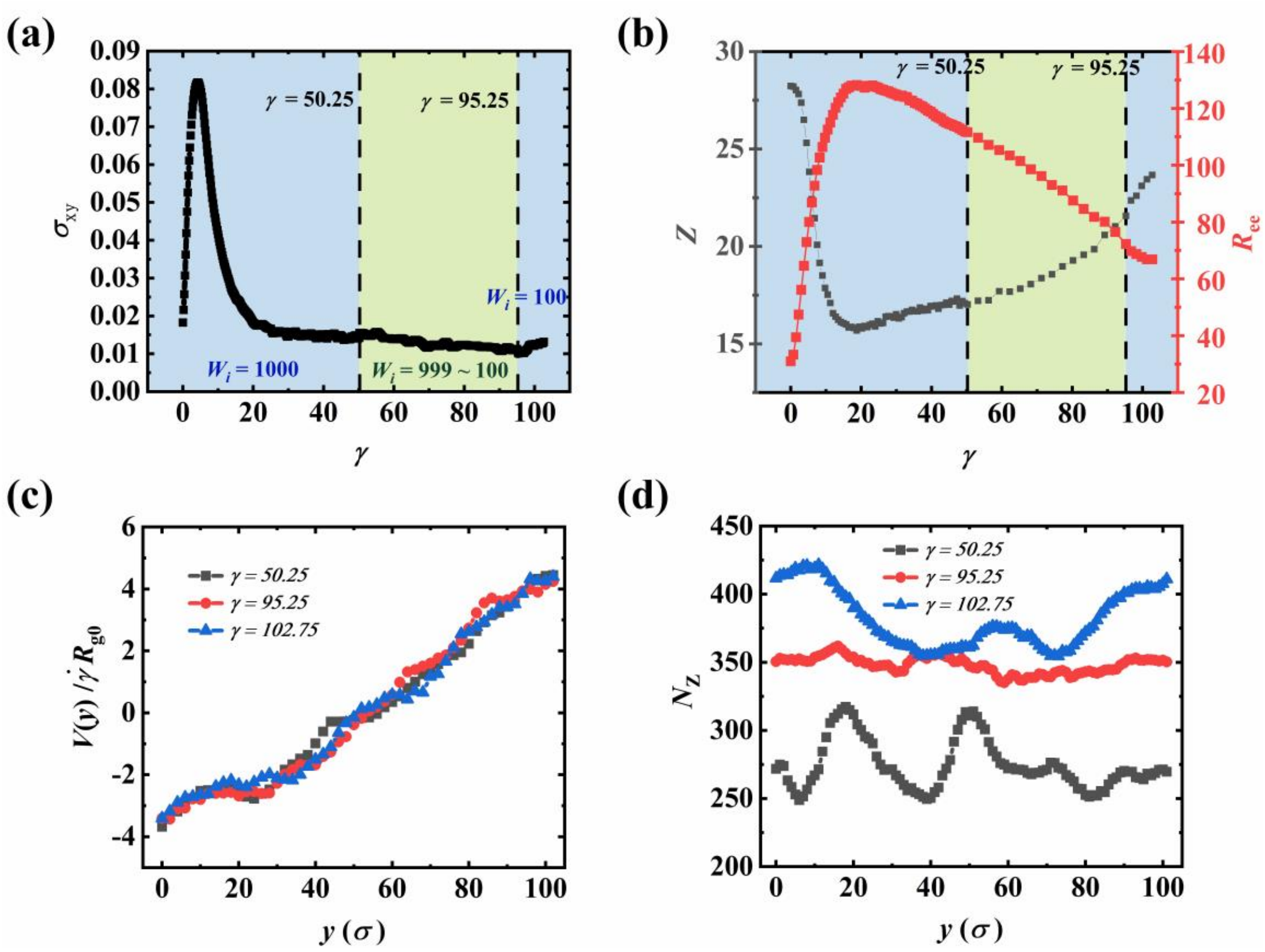

Fig. S13. (a) The shear stress as the functions of $\gamma$. In the blue-filled areas, the system is sheared at constant $W_i$ = 100 and 1000. In the red-filled area of $\gamma$ from 50.25 to 95.25, $W_i$ is gradually decreasing from 999 to 100. (b) The number of entanglement points per chain (gray symbols) and mean end-to-end distance of chains (red symbols) as the functions of $\gamma$. (c) The velocity profile at $\gamma$ = 50.25, 95.25, and 102.75. (d) The number of entanglement points in different layers at $\gamma$ = 50.25, 95.25, and 102.75.

**Table S1. The strain and stress when $Z$-$R_{ee}$ first enters the linear region and overshoot occurs**

| $W_i$ | The strain when the $Z$-$R_{ee}$ curve first enters the linear region: $\gamma_{linear}$ | The strain when overshoot occurs: $\gamma_{overshoot}$ | The stress when the $Z$-$R_{ee}$ curve first enters the linear region | The stress when overshoot occurs |
|---|---|---|---|---|
| 5 | 1.2 | 2.1 | 0.0131 | 0.0143 |
| 25 | 1.5 | 2 | 0.0228 | 0.0243 |
| 50 | 1.5 | 2 | 0.0273 | 0.0297 |
| 75 | 1.5 | 2.2 | 0.0307 | 0.0329 |
| 100 | 1.5 | 2.3 | 0.0322 | 0.0356 |
| 250 | 2 | 2.8 | 0.0450 | 0.0482 |
| 500 | 2.25 | 3.3 | 0.0562 | 0.0600 |
| 1000 | 3 | 4.25 | 0.0771 | 0.0817 |

**Table S2. The data points used for fitting *Z*-$R_{ee}$**

| $W_i$ =5 | | | $W_i$ =25 | | |
|---|---|---|---|---|---|
| *Z* | $R_{ee}$ | Corresponding strain | *Z* | $R_{ee}$ | Corresponding strain |
| 26.193 | 31 | 0 | 24.202 | 36.494 | 1.5 |
| 24.549 | 32.357 | 0.75 | 23.145 | 41.694 | 2.25 |
| 24.511 | 34.623 | 1.5 | 22.707 | 45.902 | 3 |
| 24.153 | 37.153 | 2.25 | 21.877 | 49.61 | 3.75 |
| 24.1 | 38.085 | 3 | 21.644 | 51.605 | 4.5 |
| 23.874 | 39.667 | 3.75 | 21.769 | 53.619 | 5.25 |
| 23.733 | 39.482 | 4.5 | 21.421 | 54.776 | 6 |
| 23.846 | 39.093 | 5.25 | 21.657 | 56.15 | 6.75 |
| 23.781 | 39.868 | 6 | 21.847 | 56.583 | 7.5 |
| 23.829 | 40.07763 | 9.75 | 21.774 | 57.661 | 8.25 |
| 23.819 | 39.88869 | 10.5 | 21.606 | 57.12 | 9 |
| 23.852 | 39.74477 | 11.25 | 21.426 | 57.296 | 9.75 |
| | | | 22.072 | 57.137 | 10.5 |
| | | | 21.859 | 56.771 | 11.25 |
| | | | 21.739 | 55.815 | 12 |
| | | | 21.842 | 55.244 | 12.75 |
| | | | 21.818 | 54.869 | 13.5 |
| | | | 21.988 | 55.56 | 14.25 |
| | | | 21.953 | 55.522 | 15 |
| | | | | | |
| $W_i$ =50 | | | $W_i$ =75 | | |
| *Z* | $R_{ee}$ | Corresponding strain | *Z* | $R_{ee}$ | Corresponding strain |
| 23.905 | 37.158 | 1.5 fast band | 25.709 | 32.445 | 0.75 fast band |
| 23.243 | 43.788 | 2.25 fast band | 24.684 | 36.768 | 1.5 fast band |
| 24.209 | 36.496 | 1.5 slow band | 23.849 | 42.510 | 2.25 fast band |
| 23.496 | 41.890 | 2.25 slow band | 22.531 | 49.000 | 3 fast band |
| 20.184 | 71.302 | 21 fast band | 21.749 | 58.786 | 21 fast band |
| 19.474 | 71.336 | 21.75 fast band | 21.509 | 58.952 | 21.75 fast band |
| 20.088 | 72.461 | 22.5 fast band | 21.711 | 59.260 | 22.5 fast band |
| 19.829 | 71.927 | 23.25 fast band | 21.608 | 58.171 | 23.25 fast band |
| 20.041 | 71.543 | 24 fast band | 21.903 | 58.288 | 24 fast band |
| 20.256 | 69.525 | 24.75 fast band | 21.779 | 57.583 | 24.75 fast band |
| 21.873 | 54.746 | 21 slow band | 25.545 | 33.244 | 0.75 slow band |
| 22.262 | 54.103 | 21.75 slow band | 24.782 | 37.646 | 1.5 slow band |
| 22.049 | 53.663 | 22.5 slow band | 23.998 | 43.477 | 2.25 slow band |
| 22.151 | 53.656 | 23.25 slow band | 22.861 | 48.642 | 3 slow band |
| 22.060 | 53.830 | 24 slow band | 19.413 | 75.273 | 21 slow band |
| 22.213 | 53.310 | 24.75 slow band | 19.126 | 74.189 | 21.75 slow band |
| | | | 19.709 | 74.214 | 22.5 slow band |
| | | | 19.309 | 73.958 | 23.25 slow band |
| | | | 19.305 | 75.002 | 24 slow band |
| | | | 19.462 | 74.188 | 24.75 slow band |
| | | | | | |
| **$W_i$ =100** | | | **$W_i$ =250** | | |
| ***Z*** | **$R_{ee}$** | **Corresponding strain** | ***Z*** | **$R_{ee}$** | **Corresponding strain** |
| 25.919 | 29.939 | 0 fast band | 25.000 | 45.088 | 2.25 fast band |
| 25.659 | 31.931 | 0.75 fast band | 23.903 | 52.589 | 3 fast band |
| 25.395 | 37.333 | 1.5 fast band | 22.440 | 59.839 | 3.75 fast band |
| 24.085 | 43.963 | 2.25 fast band | 21.009 | 66.392 | 4.5 fast band |
| 23.020 | 50.965 | 3 fast band | 19.380 | 73.373 | 5.25 fast band |
| 26.316 | 31.456 | 0 slow band | 17.394 | 88.074 | 20.25 fast band |
| 25.226 | 33.352 | 0.75 slow band | 17.551 | 86.868 | 21 fast band |
| 24.801 | 37.905 | 1.5 slow band | 17.711 | 86.778 | 21.75 fast band |
| 23.782 | 43.807 | 2.25 slow band | 17.827 | 85.849 | 22.5 fast band |
| 23.166 | 49.396 | 3 slow band | 17.743 | 85.209 | 23.25 fast band |
| 19.096 | 77.272 | 22.5 fast band | 17.732 | 85.290 | 24 fast band |
| 18.611 | 76.974 | 23.25 fast band | 17.917 | 85.199 | 24.75 fast band |
| 18.611 | 76.377 | 24 fast band | 25.351 | 45.585 | 2.25 slow band |
| 19.141 | 75.935 | 24.75 fast band | 24.174 | 53.295 | 3 slow band |
| 21.159 | 62.889 | 22.5 slow band | 22.753 | 60.107 | 3.75 slow band |
| 20.967 | 62.278 | 23.25 slow band | 21.445 | 66.188 | 4.5 slow band |
| 20.967 | 62.332 | 24 slow band | 20.462 | 69.107 | 5.25 slow band |
| 21.462 | 61.899 | 24.75 slow band | 19.710 | 72.281 | 20.25 slow band |
| | | | 18.845 | 78.740 | 21 slow band |
| | | | 18.904 | 79.247 | 21.75 slow band |
| | | | 18.570 | 79.162 | 22.5 slow band |
| | | | 19.057 | 79.433 | 23.25 slow band |
| | | | 19.018 | 78.811 | 24 slow band |
| | | | 18.958 | 78.674 | 24.75 slow band |
| | | | 19.250 | 78.491 | 2.25 slow band |
| | | | | | |
| **$W_i$ =500** | | | **$W_i$ =1000** | | |
| ***Z*** | ***$R_{ee}$*** | **Corresponding strain** | ***Z*** | ***$R_{ee}$*** | **Corresponding strain** |
| 24.617 | 55.338 | 3 fast band | 25.197 | 31.000 | 3 |
| 23.101 | 64.482 | 3.75 fast band | 24.292 | 33.346 | 3.75 |
| 21.631 | 72.445 | 4.5 fast band | 22.95 | 39.451 | 4.5 |
| 20.450 | 79.141 | 5.25 fast band | 21.297 | 47.459 | 5.25 |
| 19.279 | 84.530 | 6 fast band | 19.796 | 56.141 | 6 |
| 17.868 | 90.070 | 6.75 fast band | 18.742 | 64.621 | 6.75 |
| 16.801 | 94.816 | 7.5 fast band | 17.28 | 72.937 | 7.5 |
| 16.145 | 98.881 | 8.25 fast band | 16.383 | 80.035 | 8.25 |
| 15.544 | 102.163 | 9 fast band | 15.516 | 86.919 | 9 |
| 15.230 | 105.463 | 9.75 fast band | 14.818 | 92.801 | 9.75 |
| 14.927 | 108.805 | 10.5 fast band | 14.212 | 98.329 | 10.5 |
| 14.555 | 112.147 | 11.25 fast band | 14.072 | 102.731 | 11.25 |
| 14.812 | 115.164 | 12 fast band | 13.608 | 106.420 | 12 |
| 14.423 | 116.687 | 12.75 fast band | 13.445 | 109.659 | 12.75 |
| 14.305 | 117.996 | 13.5 fast band | 13.109 | 112.525 | 13.5 |
| 14.219 | 118.382 | 14.25 fast band | 12.944 | 115.073 | 14.25 |
| 14.341 | 118.042 | 15 fast band | 12.844 | 117.729 | 15 |
| 14.148 | 118.992 | 15.75 fast band | 12.851 | 120.207 | 15.75 |
| 14.313 | 119.632 | 16.5 fast band | 12.68 | 122.150 | 16.5 |
| 14.365 | 119.304 | 17.25 fast band | 12.747 | 123.820 | 17.25 |
| 14.136 | 118.915 | 18 fast band | 12.67 | 125.173 | 18 |
| 14.467 | 118.944 | 18.75 fast band | 12.68 | 126.853 | 18.75 |
| 14.277 | 119.116 | 19.5 fast band | 12.88 | 127.397 | 19.5 |
| 14.419 | 118.137 | 20.25 fast band | 12.911 | 127.984 | 20.25 |
| 14.240 | 117.574 | 21 fast band | 12.845 | 128.012 | 21 |
| 14.412 | 116.587 | 21.75 fast band | 12.727 | 128.243 | 21.75 |
| 14.448 | 115.502 | 22.5 fast band | 13.033 | 127.919 | 22.5 |
| 14.786 | 115.129 | 23.25 fast band | 12.972 | 127.874 | 23.25 |
| 14.768 | 114.661 | 24 fast band | 12.914 | 127.279 | 24 |
| 14.646 | 113.944 | 24.75 fast band | 13.096 | 127.719 | 24.75 |
| 14.660 | 112.765 | 25.5 fast band | 13.368 | 127.872 | 27 |
| 24.733 | 51.988 | 3 slow band | 13.614 | 128.080 | 30 |
| 23.630 | 57.650 | 3.75 slow band | 13.747 | 127.633 | 33 |
| 22.464 | 64.479 | 4.5 slow band | 13.913 | 127.376 | 36 |
| 21.058 | 69.932 | 5.25 slow band | 13.976 | 126.183 | 39 |
| 20.152 | 76.826 | 6 slow band | 13.936 | 124.840 | 42 |
| 19.023 | 80.776 | 6.75 slow band | 14.214 | 123.975 | 45 |
| 17.978 | 84.735 | 7.5 slow band | 14.235 | 121.644 | 48 |
| 17.741 | 88.567 | 8.25 slow band | 14.33 | 119.258 | 50.25 |
| 16.934 | 92.481 | 9 slow band | | | |
| 16.802 | 94.320 | 9.75 slow band | | | |
| 16.637 | 94.815 | 10.5 slow band | | | |
| 16.334 | 95.461 | 11.25 slow band | | | |
| 16.171 | 94.793 | 12 slow band | | | |
| 16.042 | 94.540 | 12.75 slow band | | | |
| 16.125 | 94.622 | 13.5 slow band | | | |
| 16.565 | 94.989 | 14.25 slow band | | | |
| 16.527 | 95.773 | 15 slow band | | | |
| 16.538 | 95.554 | 15.75 slow band | | | |
| 16.437 | 96.167 | 16.5 slow band | | | |
| 16.597 | 95.084 | 17.25 slow band | | | |
| 16.521 | 94.413 | 18 slow band | | | |
| 16.667 | 94.343 | 18.75 slow band | | | |
| 16.240 | 94.813 | 19.5 slow band | | | |
| 16.745 | 94.727 | 20.25 slow band | | | |
| 17.061 | 94.136 | 21 slow band | | | |
| 16.770 | 93.352 | 21.75 slow band | | | |
| 16.928 | 94.222 | 22.5 slow band | | | |
| 16.939 | 93.338 | 23.25 slow band | | | |
| 16.584 | 93.401 | 24 slow band | | | |
| 16.927 | 93.437 | 24.75 slow band | | | |
| 17.106 | 92.835 | 25.5 slow band | | | |